\documentclass{LMCS}

\usepackage{enumerate}
\usepackage{hyperref}
\usepackage{stmaryrd}
\usepackage{verbatim}
\usepackage{fancyvrb}
\usepackage{mathabx}

\usepackage[all]{hypcap}
\hypersetup{
            pdftitle=Amortised Resource Analysis with Separation Logic,
            pdfauthor=Robert Atkey}

\usepackage{mathpartir}
\usepackage{tikz}
\usetikzlibrary{shapes.multipart}

\newcommand{\link}[2]{
  \draw [fill=black] (#1.second) circle (0.3mm);
  \draw [->] (#1.second) .. controls +(right:5mm) and +(left:5mm) .. (#2.text west)  ;
}

\newcommand{\linkback}[2]{
  \draw [fill=black] (#1.second) circle (0.3mm);
  \draw [->] (#1.second) .. controls +(left:5mm) and +(right:5mm) .. (#2.text east)  ;
}

\newcommand{\linkfromtop}[2]{
  \draw [fill=black] (#1.text) circle (0.3mm);
  \draw [->] (#1.text) .. controls +(right:5mm) and +(left:5mm) .. (#2.text west)  ;
}

\newcommand{\linknull}[2]{
  \draw [fill=black] (#1.second) circle (0.3mm);
  \draw [->] (#1.second) .. controls +(right:3mm) and +(left:3mm) .. (null.west);
}

\tikzset{listnode/.style={
    rectangle split,
    rectangle split parts=2,
    draw,font=\ttfamily,
    rounded corners=1mm,
    fill=blue!10
  }
}

\tikzset{listnode2/.style={
    rectangle split,
    rectangle split parts=2,
    draw,font=\ttfamily,
    rounded corners=1mm,
    fill=green!10
  }
}

\tikzset{pairnode/.style={
    rectangle split,
    rectangle split parts=2,
    draw,
    rounded corners=1mm,
    fill=blue!10
  }
}

\tikzset{heap/.style={
    rounded corners=2mm,
    dashed,
  }
}

\def\wand{\mathrel{\textrm{---}\mkern-6mu\ast}\joinrel}

\newcommand{\pointsto}[3]{[ #1 \stackrel{\mathsf{#2}}\mapsto #3 ]}
\newcommand{\sepbar}{\mathrel|}
\newcommand{\emp}{\mathsf{emp}}

\newcommand{\dom}{\mathit{dom}}
\newcommand{\predicate}[1]{\mathsf{#1}}
\newcommand{\outputsep}{\mathop\setminus}

\def\doi{7 (2:17) 2011}
\lmcsheading%
{\doi}
{1--33}
{}
{}
{Jun.~15, 2010}
{Jun.~21, 2011}
{}   

\begin{document}

\title{Amortised Resource Analysis with Separation Logic}

\author[R.~Atkey]{Robert Atkey}
\address{University of Strathclyde, UK}
\email{Robert.Atkey@cis.strath.ac.uk}


\begin{abstract}
  Type-based amortised resource analysis following Hofmann and
  Jost---where resources are associated with individual elements of
  data structures and doled out to the programmer under a linear
  typing discipline---have been successful in providing concrete
  resource bounds for functional programs, with good support for
  inference. In this work we translate the idea of amortised resource
  analysis to imperative pointer-manipulating languages by embedding a
  logic of resources, based on the affine intuitionistic Logic of
  Bunched Implications, within Separation Logic. The Separation Logic
  component allows us to assert the presence and shape of mutable data
  structures on the heap, while the resource component allows us to
  state the consumable resources associated with each member of the
  structure.

  We present the logic on a small imperative language, based on Java
  bytecode, with procedures and mutable heap. We have formalised the
  logic and its soundness property within the Coq proof assistant and
  extracted a certified verification condition generator. We also
  describe an proof search procedure that allows generated
  verification conditions to be discharged while using linear
  programming to infer consumable resource annotations.

  We demonstrate the logic on some examples, including proving the
  termination of in-place list reversal on lists with cyclic tails.
\end{abstract}

\maketitle

\section{Introduction}\label{sec:intro}

Tarjan, in his paper introducing the concept of amortised complexity
analysis \cite{tarjan85}, noted that the statement and proof of
complexity bounds for operations on some data structures can be
simplified if we think of the data structure as being able to store
``credits'' that are used up by later operations. By setting aside
credit inside a data structure to be used by later operations, the cost
of a sequence of operations can be amortised over time. In this paper,
we propose a way to merge amortised complexity analysis with
Separation Logic \cite{ishtiaq01biassertion,reynolds02separation} to
formalise some of these arguments and to simplify the specification
and verification of resource-consuming pointer-manipulating programs.

Separation Logic is built upon a notion of resources and their
separation. The assertion $A * B$ holds for some resource if it can be
split into two separate resources that make $A$ true and $B$ true
respectively. Resource separation enables local reasoning about
mutation of resources; if the program mutates the resource associated
with $A$, then we know that $B$ is still true on its separate
resource. Usually Separation Logic uses mutable heaps as its notion
of resource. In this paper, we combine heaps with consumable resources
to reason about resource consumption as well as resource mutation.

To see how Separation Logic-style reasoning about consumable resources
is beneficial, consider the standard inductively defined list segment
predicate from Separation Logic, augmented with an additional
proposition $R$ denoting the presence of a consumable resource for
every element of the list:
\begin{displaymath}
  \predicate{lseg}(R, x,y) \equiv
  \begin{array}[t]{l}
    \hspace{0.3cm} x = y \land \predicate{emp} \\
    \mathop\lor \exists d, z.\ \pointsto{x}{data}{d} * \pointsto{x}{next}{z} * R * \predicate{lseg}(z,y)
  \end{array}
\end{displaymath}
We will introduce the assertion logic properly in
\autoref{sec:assertion-logic} below. We can represent a heap $H$ and a
consumable resource $r$ that satisfy this predicate graphically:
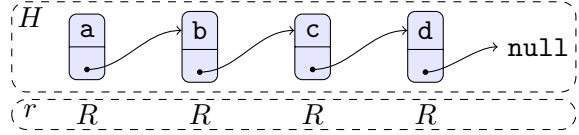
\begin{center}
  \begin{tikzpicture}
    \draw [heap] (-1,0.6) rectangle (6.5,-0.6);
    \node at (-0.75,0.4) {$H$};
    \draw [heap] (-1,-0.7) rectangle (6.5,-1.1);
    \node at (-0.75,-.85) {$r$};

    \node (null) [font=\ttfamily] at (6,0) {null};

    \foreach \n / \x in {a/0, b/1.5, c/3, d/4.5} {
        \node (\n) [listnode] at (\x,0) {\n};
        \node at (\x,-.9) {$R$};
    }

    \foreach \a / \b in {a/b, b/c, c/d} \link{\a}{\b};
    \linknull{d}

  \end{tikzpicture}
\end{center}
So we have $r, H \models \predicate{lseg}(R, x,\mathit{null})$,
assuming $x$ contains the address of the head of the list. Here $r = R
\cdot R \cdot R \cdot R$---we assume that consumable resources form a
commutative monoid---and $r$ represents the resource that is available
for the program to use in the future. We can split $H$ and $r$ to
separate out the head of the list with its associated resource:
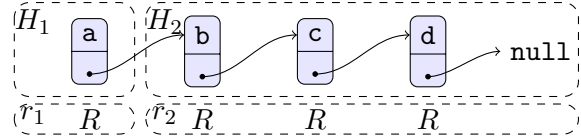
\begin{center}
  \begin{tikzpicture}
    \draw [heap] (-1,0.65) rectangle (0.6,-0.6);
    \node at (-0.75,0.4) {$H_1$};
    \draw [heap] (-1,-0.7) rectangle (0.6,-1.1);
    \node at (-0.75,-.85) {$r_1$};
    \draw [heap] (0.75,0.65) rectangle (6.5,-0.6);
    \node at (1,0.4) {$H_2$};
    \draw [heap] (0.75,-0.7) rectangle (6.5,-1.1);
    \node at (1,-.85) {$r_2$};

    \node (null) [font=\ttfamily] at (6,0) {null};

    \foreach \n / \x in {a/0, b/1.5, c/3, d/4.5} {
        \node (\n) [listnode] at (\x,0) {\n};
        \node at (\x,-.9) {$R$};
    }

    \foreach \a / \b in {a/b, b/c, c/d} \link{\a}{\b};
    \linknull{d}

  \end{tikzpicture}
\end{center}
This heap and resource satisfy $r_1 \cdot r_2, H_1 \uplus H_2 \models
\pointsto{x}{data}{\texttt{a}} * \pointsto{x}{next}{y} * R *
\predicate{lseg}(R, y,\mathit{null})$, where $H_1 \uplus H_2 = H$,
$r_1 \cdot r_2 = r$ and we assume that $y$ contains the address of the
\texttt{b} element. Now that we have separated out the head of the
list and its associated consumable resource, we are free to mutate the
heap $H_1$ and consume the resource $r_1$ without affecting the tail
of the list. So the program can move to a new state where the head of
the list has been mutated to \texttt{A} and the associated resource
has been consumed:
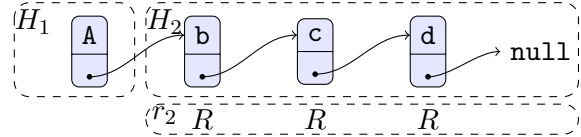
\begin{center}
  \begin{tikzpicture}
    \draw [heap] (-1,0.65) rectangle (0.6,-0.6);
    \node at (-0.75,0.4) {$H_1$};
    \draw [heap] (0.75,0.65) rectangle (6.5,-0.6);
    \node at (1,0.4) {$H_2$};
    \draw [heap] (0.75,-0.7) rectangle (6.5,-1.1);
    \node at (1,-.85) {$r_2$};

    \node (null) [font=\ttfamily] at (6,0) {null};

    \foreach \n / \x in {A/0, b/1.5, c/3, d/4.5} {
        \node (\n) [listnode] at (\x,0) {\n};
        \ifnum 0<\x \node at (\x,-.9) {$R$}; \fi
    }

    \foreach \a / \b in {A/b, b/c, c/d} \link{\a}{\b};
    \linknull{d}

  \end{tikzpicture}
\end{center}
We do not need to do anything special to reason that the tail of the
list and its associated consumable resource are unaffected.

The combined assertion about heap and consumable resource describes
the current shape and contents of the heap and also the available
resource that the program may consume in the future. By ensuring that,
for every state in the program's execution, the resource consumed plus
the resource available for consumption in the future is less than or
equal to a predefined bound, we can ensure that the entire execution
is resource bounded. This is the main assertion of soundness for our
program logic in \autoref{sec:program-logic-soundness}. We also treat
a variant program logic that allows dynamic but fallible resource
acquisition in \autoref{sec:acquisition}.

By intermixing resource assertions with Separation Logic assertions
about the shapes of data structures, as we have done with the resource
carrying $\predicate{lseg}$ predicate above, we can specify amounts of
resource that depend on the shape of data structures in memory. By the
definition of $\predicate{lseg}$, we know that the amount of resource
available to the program is linearly proportional to the length of the
list, without having to do any arithmetic reasoning about lengths of
lists. In \autoref{sec:inductive-preds} we describe some useful
inductively defined predicates that maintain a close connection
between shape and resources.

The association of resources with parts of a data structure is exactly
the banker's approach to amortised complexity analysis proposed by
Tarjan.

Our original inspiration for this work came from the work of Hofmann
and Jost \cite{hofmann-jost} on the automatic heap-space analysis of
functional programs. Their analysis associates with every element of a
data structure a permission to use a piece of resource (in their case,
heap space). This resource is made available to the program when the
data structure is decomposed using pattern matching. When constructing
part of a data structure, the required resources must be available. A
linear type system is used to ensure that data structures carrying
resources are not duplicated since this would entail duplication of
consumable resource. This scheme was later extended to imperative
object-oriented languages \cite{raja1,raja2}, but still using a
type-based analysis.

\subsection{Contributions} We summarise the content and contributions
of this work:
\begin{enumerate}[$\bullet$]
\item In \autoref{sec:logic}, we define a program logic that allows
  mixing of assertions about heap shapes and assertions about future
  consumable resources. Tying these together allows us to easily state
  resource properties in terms of the shapes of heap-based data
  structures, rather than in terms of extensional properties such as
  their size or contents. We have formalised the soundness proof of
  our program logic in the Coq proof assistant.
\item In \autoref{sec:assertion-logic} we present a syntax for our
  assertion logic, based on a combination of Boolean Bunched
  Implications extended with pointer assertion primitives, as is usual
  for Separation Logic, and affine intuitionistic Bunched Implications
  to declaratively state properties of the consumable resource
  available to a program. We also discuss several useful inductively
  defined predicates for this logic that demonstrate tight connections
  between heap shapes and resources. We also discuss the advantages
  and disadvantages of this tight connection.
\item In \autoref{sec:proofsearch}, we define a restricted subset of
  the assertion logic that allows us to perform effective proof search
  to discharge verification conditions, while inferring resource
  annotations. A particular feature of the proof search procedure we
  describe is that, given loop invariants that specify only the the
  shape of data structures, we can infer the necessary consumable
  resource annotations.
\item In \autoref{sec:examples} and \autoref{sec:auto-examples}, we
  demonstrate the logic on some examples, showing how a mixture of
  amortised resource analysis and Separation Logic can be used to
  simplify resource-aware specifications, deal with relatively complex
  pointer manipulation and to prove termination in the presence of
  cyclic structures in the heap.
\end{enumerate}

\subsection{Differences to previously published versions}

This paper is a revised and expanded version of the ESOP 2010
conference version \cite{atkey10amortised}. Additional explanation has
been provided and \autoref{sec:acquisition} has been added on a
variant program logic accounting for dynamic resource
acquisition. \autoref{sec:assertion-logic} has been expanded with more
examples of inductively defined predicates tightly integrating shape
and resource properties, and a discussion of the disadvantages of this
integration. \autoref{sec:proofsearch} on automated verification and
proof search has been refined for clarity and to better match the
implementation. We have also included the merge-sort example in
\autoref{sec:merge-sort} from an invited TGC 2010 paper
\cite{aspinall10tgc}.

\section{Motivating Examples}\label{sec:examples}

The example we gave in the \hyperref[sec:intro]{introduction}, where a
program iterates through a list consuming resources as it proceeds,
only demonstrates an extremely simple, albeit common, pattern. In this
section, we give two more complex examples that serve to highlight the
advantages of the amortised approach to specifying and verifying
resource bounds. The first example, in \autoref{sec:func-queues},
demonstrates how the integrated description of available consumable
resources and heap structure helps with specification. The second
example, in \autoref{sec:merge-sort}, shows how the amortised approach
simplifies the problem of verifying the resource consumption of a
pointer manipulating program. This example also shows a weakness of
maintaining a tight connection between heap shape and consumable
resources, which we discuss further in
\autoref{sec:resource-separation}.

In each case we attempt to demonstrate how amortised reasoning is
easier than the traditional approach of keeping a global counter for
consumed resources as a ``ghost'' variable in the logic.

\subsection{Functional Queues}\label{sec:func-queues}

We consider so-called functional queues
\cite{hood-melville81,burton82}, where a queue is represented by a
pair of lists. This example is a standard one for introducing
amortised complexity analysis \cite{okasaki98}. We verify an
imperative implementation that performs mutations in-place on the
underlying lists. The point of this example is to see how the
amortised technique simplifies the specifications of the procedures
operating on this data structure.
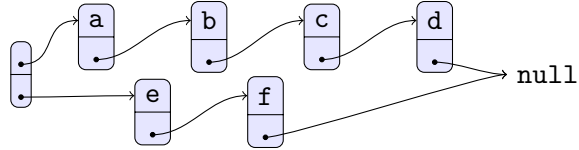
\begin{center}
  \begin{tikzpicture}
    \node (null) [font=\ttfamily] at (7,0) {null};

    \node (control) [pairnode] at (0,0) {\ };

    \foreach \n / \x in {a/0, b/1, c/2, d/3} {
      \node (\n) [listnode] at (\x*1.5 + 1, 0.5) {\n};
    }

    \foreach \n / \x in {e/0, f/1} {
      \node (\n) [listnode] at (\x*1.5 + 1.75, -0.5) {\n};
    }

    \foreach \a / \b in {a/b, b/c, c/d, e/f} \link{\a}{\b};
    \linknull{d}; \linknull{f};

    \link{control}{e}
    \linkfromtop{control}{a}

  \end{tikzpicture}
\end{center}

The top list represents the head of the queue, while the bottom list
represents the tail of the queue in reverse. This structure represents
the queue
$[\texttt{a},\texttt{b},\texttt{c},\texttt{d},\texttt{f},\texttt{e}]$. When
we enqueue a new element, we add it to the head of the bottom list. To
dequeue an element, we remove it from the head of the top list. If the
top list is empty, then we reverse the bottom list and change the top
pointer to point to it, changing the bottom pointer to point to
$\mathit{null}$, representing the empty list.

When determining the complexity of these operations, it is obvious
that the enqueue operation is constant time, but the dequeue operation
either takes constant time if the top list is empty, or takes time
linear in the size of the bottom list in order to perform the
reversal. If we were to account for resource usage by maintaining a
global counter then we would have to expose the lengths of the two
lists in specification of the enqueue and dequeue instructions. So we
would need a predicate $\predicate{queue}(x, h, t)$ to state that $x$
points to a queue with a head and tail lists of lengths $h$ and $t$
respectively. The operations would have the specifications (written as
Hoare triples):
\begin{displaymath}
  \begin{array}{rrcl}
  \forall r_1. & \{\predicate{queue}(x,h,t) \land r_c = r_1\} & \texttt{enqueue} & \{\predicate{queue}(x,h,t+1) \land r_c = r_1 + R\} \\
  \forall r_1. & \{\predicate{queue}(x,0,t) \land r_c = r_1\} & \texttt{dequeue} & \{\predicate{queue}(x,t\dotdiv 1,0) \land r_c = r_1 + (1 + t)R\} \\
  \forall r_1. & \{\predicate{queue}(x,h+1,t) \land r_c = r_1\} & \texttt{dequeue} & \{\predicate{queue}(x,h,t) \land r_c = r_1 + R\}
  \end{array}
\end{displaymath}
where $r_c$ is a ghost variable counting the total amount of resource
consumed by the program, and $R$ is the amount of resource required to
perform a single list node manipulation. The dotted minus $t \dotdiv
1$ denotes the predecessor operation on natural numbers. Note that we
have had to give two specifications for \texttt{dequeue} for the cases
when the head list is empty and when the head list has an element. The
accounting for the sizes of the internals of the queue data structure
is of no interest to clients of this data structure. These
specifications complicate clients' reasoning when using these queues.

Using amortised analysis, this specification can be drastically
simplified. We associate a single piece of resource with each element
of the tail list so that when we come to reverse the list we have the
necessary resource available to reverse each list
element. The queue predicate is therefore:
\begin{displaymath}
  \predicate{queue}(x) \equiv \exists y, z.\ \pointsto{x}{head}{y} * \pointsto{x}{tail}{z} * \predicate{lseg}(e, y,\mathit{null}) * \predicate{lseg}(R, z,\mathit{null})
\end{displaymath}
where $\predicate{lseg}$ is the resource-carrying list predicate given
above and $e$ is the unit of the consumable resource monoid,
representing no resource. The list holding the head of the queue has
no resource associated with every element, while the list holding the
tail of the queue does have a resource associated with every element,
waiting to pay for the list reversal when it occurs. Diagrammatically,
a queue with associated resources looks like this:
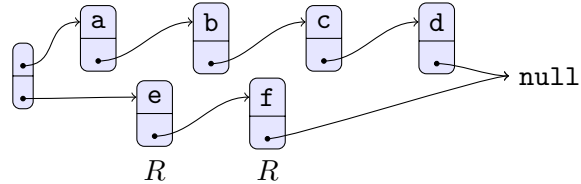
\begin{center}
  \begin{tikzpicture}
    \node (null) [font=\ttfamily] at (7,0) {null};

    \node (control) [pairnode] at (0,0) {\ };

    \foreach \n / \x in {a/0, b/1, c/2, d/3} {
      \node (\n) [listnode] at (\x*1.5 + 1, 0.5) {\n};
    }

    \foreach \n / \x in {e/0, f/1} {
      \node (\n) [listnode] at (\x*1.5 + 1.75, -0.5) {\n};
      \node at (\x*1.5 + 1.75, -1.25) {$R$};
    }

    \foreach \a / \b in {a/b, b/c, c/d, e/f} \link{\a}{\b};
    \linknull{d}; \linknull{f};

    \link{control}{e}
    \linkfromtop{control}{a}

  \end{tikzpicture}
\end{center}

The specifications of the operations now becomes straightforward:
\begin{mathpar}
  \{\predicate{queue}(x) * R * R\}\texttt{enqueue}\{\predicate{queue}(x)\}

  \{\predicate{queue}(x) * R\}\texttt{dequeue}\{\predicate{queue}(x)\}
\end{mathpar}
To enqueue an element, we require two elements of resource: one to add
the new element to the tail list, and one to ``store'' in the list so
that we may use it for a future reversal operation. To dequeue an
element, we require a single element of resource to remove an element
from a list. If a list reversal is required then it is paid for by the
resources previously required by $\texttt{enqueue}$.

Once we have set the specification of queues to store one element of
resource for every node in the tail list, we can use the resource
annotation inference procedure presented in \autoref{sec:proofsearch}
to generate the resource parts of the \texttt{enqueue} and
\texttt{dequeue} specifications.

\subsection{Merge-sort Inner Loop}\label{sec:merge-sort}

We now describe a more complicated list manipulating program that
shows the benefits of the amortised approach for verification. This
example demonstrates the combination of reasonably complex pointer
manipulation with resource reasoning. Most of the technical details
arise from dealing with the heap-shape behaviour of the program; the
resource bounds simply drop out thanks to the inference of resource
annotations.

\begin{figure}
  \centering
\begin{Verbatim}[commandchars=@??]
public static Node mergeInner (Node list, int k) {
     Node p      = list;
     Node tail   = null;

     list = null;

     while (p != null) {
          Node q = p;
          for (int i = 0; i < k; i++) {
   	    q = q.next;
   	    if (q == null) break;
          }

          Node pstop = q;
          int qsize = k;
          while (p != pstop || (qsize > 0 && q != null)) {
   	    Node e;
   	    if (p == pstop) {
   		 e = q;
   		 q = q.next;
   		 qsize--;
   	    } else if (qsize == 0 || q == null) {
   		 e = p;
   		 p = p.next;
   	    } else if (p.data <= q.data) {
   		 e = p;
   		 p = p.next;
   	    } else {
   		 e = q;        // perform swap
   		 q = q.next;
   		 qsize--;
   	    }
   	    
   	    if (tail != null)  tail.next = e;
   	    else               list = e;
   	    
   	    tail = e;
          }

          p = q;
     }
     
     if (tail == null) return null; else tail.next = null;
     return list;
}
\end{Verbatim}
  \caption{Inner loop of an in-place linked-list merge sort}
  \label{fig:merge-sort-code}
\end{figure}

\DefineShortVerb{\|}

Consider the Java method declaration |mergeInner| shown in
\autoref{fig:merge-sort-code} that describes the inner loop of an
in-place merge sort algorithm for linked lists\footnote{Adapted from C
  code:
  \url{http://www.chiark.greenend.org.uk/\~sgtatham/algorithms/listsort.html}.}. The
method takes two arguments: |list|, a reference to the head node of a
linked list; and |k|, an integer. The integer argument dictates the
sizes of the sublists that the method will be merging in this
pass. The method steps through the list |2*k| elements at a time,
merging the two sublists of length |k| each time. The outer loop does
the |2*k| stepping, and the inner loop does the merging. To accomplish
a full merge sort, |mergeInner| would be called $\log_2 n$ times with
doubling |k|, where $n$ is the length of the list.

Assume that we wish to account for the number of swapping operations
performed by this method, i.e.\ the number of times that the fourth
branch of the |if| statement in the inner loop is executed. We
accomplish this in our implementation by inserting a special |consume|
instruction at this point.

The pre- and post-conditions of the method are as follows:
\begin{eqnarray*}
  \textrm{Pre}(\texttt{mergeInner}) & : & \texttt{list} \not= \mathit{null} \land (\predicate{lseg}(x,\texttt{list},\mathit{null}) * R^y) \\
  \textrm{Post}(\texttt{mergeInner}) & : & \predicate{lseg}(0, \texttt{retval}, \mathit{null})
\end{eqnarray*}

The precondition states that the first argument points to a list
segment ending with |null|, with $x$ amount of resource associated
with every element of the list, and $y$ amount of additional resource
that may be used. The actual values of $x$ and $y$ will be inferred by
a linear program solver. The condition $\texttt{list} \not=
\mathit{null}$ is a safety condition required to prevent a null
pointer exception.

The outer loop in the method needs a disjunctive invariant
corresponding to whether this is the first iteration or a later
iteration.
\begin{displaymath}
  \begin{array}{ll}
         & (\predicate{lseg}(o_1, \texttt{list}, \texttt{tail}) * \pointsto{\texttt{tail}}{next}{?} * \pointsto{\texttt{tail}}{data}{?} * \predicate{lseg}(o_2, \texttt{p}, \mathit{null}) * R^{o_3})\\
    \lor & ((\texttt{list} = \mathit{null} \land \texttt{tail} = \mathit{null}) * \predicate{lseg}(o_4, \texttt{p}, \mathit{null}) * R^{o_5})
  \end{array}
\end{displaymath}

The first disjunct is used on normal iterations of loop: the variable
|list| points to the list that has been processed so far, ending at
|tail|; and |p| points to the remainder of the list that is to be
processed. We have annotated these lists with the resource variables
$o_1$ and $o_2$ that will contain the resources associated with each
element of these lists. The second disjunct covers the case of the
first iteration, when |list| and |tail| are null and |p| points to the
complete list to be processed. The resource annotation $o_4$ will be
filled in with the amount of resource that is required for every
element of the list to be processed. As one might expect, this will be
equal to the value of $x$, the amount of resource required by the
precondition of the method.

Moving on, we consider the first inner loop that advances the pointer
|q| by |k| elements forward, thus splitting the list ahead of |p| into
a |k|-element segment and the rest of the list. The next loop will
merge the first |k|-element segment with the |k|-element prefix of the
second segment. It is convenient for our implementation to split out
this inner loop into another method\footnote{This is because our
  implementation works on unstructured JVM-like bytecode, and so
  cannot easily apply Separation Logic's frame rule to modularly
  reason about the loop. Using a separate method allows application of
  the frame rule. See \hyperref[rem:frame-rule]{Remark
    \ref*{rem:frame-rule}}.}, with the following signature:
\begin{Verbatim}
public static Node advance (Node l, int k)
\end{Verbatim}

The argument |l| points to a linked list, and the method will advance
|k| elements through the list (or until the end) and return a pointer
to the split point. The specification of this method is:
\begin{eqnarray*}
  \textrm{Pre}(\texttt{advance}) & : & \predicate{lseg}(a_0, \texttt{l}, \mathit{null})\\
  \textrm{Post}(\texttt{advance}) & : & \predicate{lseg}(a_0, \texttt{l}, \texttt{retval}) * \predicate{lseg}(a_0, \texttt{retval}, \mathit{null})
\end{eqnarray*}

Again, we have left the resource annotation on the elements of the
list as a variable $a_0$, to be filled in by the linear solver. The
appearance of the same variable in the pre- and post-condition implies
that we expect this resource to be preserved by the method. The fact
that we have had to explicitly mention the resources that must be
preserved points to a limitation of the amortised method as we present
it here. We discuss this issue further in
\autoref{sec:resource-separation}.

Proceeding though our main method, the invariant of the inner loop is
as follows, again as two disjuncts according to whether it is the
first or later iteration of the outer loop:
\begin{displaymath}
  \begin{array}{ll}
         & (\predicate{lseg}(i_1, \texttt{list}, \texttt{tail}) * \pointsto{\texttt{tail}}{next}{?} * \pointsto{\texttt{tail}}{data}{?} \\
         & \hspace{3cm} *\ \predicate{lseg}(i_2, \texttt{p}, \texttt{pstop}) * \predicate{lseg}(i_3,\texttt{q},\mathit{null}) * R^{i_4})\\
    \lor & ((\texttt{list} = \mathit{null} \land \texttt{tail} = \mathit{null}) * \predicate{lseg}(i_5, \texttt{p}, \texttt{pstop}) * \predicate{lseg}(i_6, \texttt{q}, \mathit{null}) * R^{i_7})
  \end{array}
\end{displaymath}
The first part of each disjunct is as before, stating that |list| to
|tail| contains the part of list that has been processed. Since we
have now split the remainder of the list into two pieces we have two
separate list segments referenced by |p| and |q| pointing to the parts
of the list that are to be merged. The resource meta-variable $i_1$
will indicate the amount of resource associated with the list that has
been processed; $i_2$ and $i_5$ are the amount of resource associated
with the ``left-hand'' pending list; and $i_3$ and $i_4$ are the
resource associated with the ``right-hand'' pending list.

At the end of the method, we null terminate the list and return. A
superfluous null check on |this| is required for our tool to prove
memory safety. The fact that |this| is non-null relies on the
execution of the inner loop at least once, which requires that $k >
0$. This fact is not expressible in the logic of the implementation
described in \autoref{sec:proofsearch}.

Running this example through our implementation produces the solution
$x = 1$, $y = 0$ for the precondition's resource annotations. This
indicates that the input list needs to contain one element of resource
for every list element. For the outer loop's invariant, we obtain $o_2
= o_4 = 1$ and all the others are $0$. This indicates that the list we
have processed has had all its resources consumed, while the list
remaining to be processed still has associated resources. This is as
expected for a loop iterating through a list. The specification of
|advance| is completed by inferring $a_0 = 1$, indicating that
|advance| preserves the resources associated with the list. Finally
the inner loop's invariant has $i_2 = i_3 = i_5 = i_6 = 1$ and all
others $0$, indicating that the two list segments that are remaining
to be processed have associated resources, while the processed
segments do not.

\subsubsection{Comparisons to other techniques.}

Though we have had to work to supply the loop invariants for our
implementation, we note that these invariants may be inferred by other
tools, for example \cite{berdine05symbolic}, and the resource
variables automatically inserted using the procedure outlined in
\autoref{sec:proofsearch}. The key to the amortised approach is the
tight connection between shape invariants, which is a complex but
well-studied problem, and resource consumption.

Most other techniques for resource usage analysis that handle data
structures do so by considering the sizes of the structures. The SPEED
system of Gulwani et al. \cite{gulwani09speed} can infer resource
bounds for programs manipulating heap-based data structures, but only
when these data structures are manipulated through abstract
interfaces. The specifications for these abstract interfaces record
the effect of the operations on the size of the data structure. Thus,
the technique is unable to cope with the kind of program that we have
presented above that uses direct pointer manipulation. Nevertheless,
Gulwani et al. report impressive results on real-world Microsoft
product code.

The COSTA system \cite{albert07costa} can deal with some uses of
direct pointer manipulation, but accounts for the sizes of heap-based
data structures by counting the length of the longest path from a
given reference. Thus, it cannot deal with programs that demonstrate
sharing on the heap; the Java method described above has, in its inner
loop, three pointers all pointing to the same list.

One might also use Separation Logic to deal with sharing on the heap,
and add information on the sizes of heap-based data structures to
account for resource usage. So one would have a predicate
$\predicate{lseg}^n(x,y)$ that describes a list segment of length $n$
from $x$ to $y$, along with a ghost variable to track resource
consumption as discussed in \autoref{sec:func-queues}. We argue that
the amortised approach described here is simpler due to the
differences in reasoning between the \emph{global} property of the
length of a whole list, and the \emph{local} property of each list
element having an associated amount of resource to be used. For
example, consider the specification of the |advance| method using
sized structures:
\begin{eqnarray*}
  \textrm{Pre}(\texttt{advance}) & : & \predicate{lseg}^n(\texttt{l}, \mathit{null})\\
  \textrm{Post}(\texttt{advance}) & : & \exists n_1, n_2.\ n_1 + n_2 = n \land (\predicate{lseg}^{n_1}(\texttt{l}, \texttt{retval}) * \predicate{lseg}^{n_2}(\texttt{retval}, \mathit{null}))
\end{eqnarray*}
We have had to introduce two existential variables indicating the
sizes of the lists returned by the method. These additional values
have to then be related back to the length of the original list by the
calling method, and thence to the resource consumption, requiring
non-straightforward arithmetic reasoning. The amortised approach
exploits the shape-reasoning already present in Separation Logic to
account for resources.

\UndefineShortVerb{\|}

\section{A Program Logic for Heap and Resources}\label{sec:logic}

We now describe a simple programming language and a
consumable-resource-aware logic for it. We define a ``shallow''
program logic where we treat pre- and post-conditions and program
assertions as arbitrary predicates over heaps and consumable
resources. In \autoref{sec:assertion-logic}, we will layer on top of
this a ``deep'' assertion logic where predicates are actually
Separation Logic formulae augmented with atomic resource
propositions.

At the \hyperref[sec:acquisition]{end of this section}, we consider a
variant system that allows dynamic but fallible resource acquisition
as well as resource consumption.

The development in this section has been formalised within the Coq
proof assistant; thus the shallow embedding makes use of Coq's own
logic as the assertion logic. We also make minor use of Coq's
dependent types. \hyperref[lem:frame-safety]{Lemma
  \ref*{lem:frame-safety}} and \hyperref[thm:soundness]{Theorem
  \ref*{thm:soundness}} establishing the soundness of the program
logic have been mechanically verified within Coq, as have
\hyperref[lem:frame-safety-2]{Lemma \ref*{lem:frame-safety-2}} and
\hyperref[thm:soundness-2]{Theorem \ref*{thm:soundness-2}}
establishing soundness for the dynamic resource acquisition variant.

\subsection{Semantic Domains}

Assume an infinite set $\mathbb{A}$ of memory addresses. We model
heaps as finite partial maps $\mathbb{H} = (\mathbb{A} \times
\mathbb{F}) \rightharpoonup_{\mathit{fin}} \mathbb{V}$, where
$\mathbb{F}$ ranges over field names and $\mathbb{V} = \mathbb{A}_\bot
+ \mathbb{Z}$ represents the values that programs can directly
manipulate: possibly null addresses and integers. We write $\dom(H)$
for the domain of a heap and $H_1\#H_2$ for heaps with disjoint
domains; $H_1 \uplus H_2$ denotes union of heaps with disjoint
domains.

Consumable resources are represented as elements of an ordered monoid
$(\mathcal{R}, \sqsubseteq, \cdot, e)$, where $e$ is the least
element. Example consumable resources include $(\mathbb{N}, \leq, +,
0)$ or $(\mathbb{Q}^{\geq 0}, \leq, +, 0)$ for representing a single
resource that is consumed (e.g. time or space), or multisets for
representing multiple named resources that may be consumed
independently. The ordering on consumable resources is used to allow
weakening in our assertion logic: we allow the asserter to assert that
more resources are required by the program than are actually needed.

As is standard in the semantics of substructural logics
\cite{restall}, we make use of a ternary relation to describe the
combination of separate entities. In our case, the entities are pairs
of heaps and consumable resources:
\begin{displaymath}
  Rxyz\ \Leftrightarrow\begin{array}[t]{l}
    H_1 \# H_2 \land H_1 \uplus H_2 = H_3 \land r_1 \cdot r_2 \sqsubseteq r_3 \\
    \quad\textrm{where }x = (H_1,r_1), y = (H_2,r_2), z = (H_3,r_3)
  \end{array}
\end{displaymath}
We extend the order on resources to pairs of heaps and resources by
$(H_1,r_1) \sqsubseteq (H_2,r_2)$ iff $H_1 = H_2$ and $r_1 \sqsubseteq
r_2$.

\subsection{A Little Virtual Machine}

\newcommand{\instr}[1]{\mathsf{#1}}

\newcommand{\frmstep}{\stackrel{\mathsf{frm}}\longrightarrow}
\newcommand{\mutstep}{\stackrel{\mathsf{mut}}\longrightarrow}
\newcommand{\step}{\stackrel{\mathit{prg}}\longrightarrow}
\newcommand{\actframe}[3]{\langle \mathit{code}, #1, #2, #3 \rangle}
\newcommand{\pc}{\mathit{pc}}
\newcommand{\sem}[1]{\llbracket #1 \rrbracket}
\newcommand{\append}{\mathop{+\kern-0.5em+}}

\begin{figure}[b]
  \centering
  \begin{mathpar}
    \inferrule*
    {\mathit{code}[\pc] = \instr{iconst}\ z}
    {\actframe{S}{L}{\pc} \frmstep \actframe{z::S}{L}{\pc+1}}

    \inferrule*
    {\mathit{code}[\pc] = \instr{ibinop}\ \mathit{op}}
    {\actframe{z_1::z_2::S}{L}{\pc} \frmstep \actframe{(\sem{\mathit{op}}\ z_1\ z_2)::S}{L}{\pc+1}}

    \inferrule*
    {\mathit{code}[\pc] = \instr{pop}}
    {\actframe{z::S}{L}{\pc} \frmstep \actframe{S}{L}{\pc+1}}

    \inferrule*
    {\mathit{code}[\pc] = \instr{load}\ n \\ L[n] = v}
    {\actframe{S}{L}{\pc} \frmstep \actframe{v::S}{L}{\pc+1}}

    \inferrule*
    {\mathit{code}[\pc] = \instr{store}\ n}
    {\actframe{v::S}{L}{\pc} \frmstep \actframe{S}{L[n\mapsto v]}{\pc+1}}

    \inferrule*
    {\mathit{code}[\pc] = \instr{aconst\_null}}
    {\actframe{S}{L}{\pc} \frmstep \actframe{\mathit{null}::S}{L}{\pc+1}}

    \inferrule*
    {\mathit{code}[\pc] = \instr{binarycmp}\ \mathit{cmp}\ \mathit{offset} \\ z_1 \mathrel{\sem{\mathit{cmp}}} z_2}
    {\actframe{z_1::z_2::S}{L}{\pc} \frmstep \actframe{S}{L}{\mathit{offset}}}

    \inferrule*
    {\mathit{code}[\pc] = \instr{binarycmp}\ \mathit{cmp}\ \mathit{offset} \\  \lnot(z_1 \mathrel{\sem{\mathit{cmp}}} z_2)}
    {\actframe{z_1::z_2::S}{L}{\pc} \frmstep \actframe{S}{L}{\pc+1}}

    \inferrule*
    {\mathit{code}[\pc] = \instr{unarycmp}\ \mathit{cmp}\ \mathit{offset} \\ z \mathrel{\sem{\mathit{cmp}}} 0}
    {\actframe{z::S}{L}{\pc} \frmstep \actframe{S}{L}{\mathit{offset}}}

    \inferrule*
    {\mathit{code}[\pc] = \instr{unarycmp}\ \mathit{cmp}\ \mathit{offset} \\ \lnot(z \mathrel{\sem{\mathit{cmp}}} 0)}
    {\actframe{z::S}{L}{\pc} \frmstep \actframe{S}{L}{\pc+1}}

    \inferrule*
    {\mathit{code}[\pc] = \instr{ifnull}\ \mathit{offset} \\ a = \mathit{null}}
    {\actframe{a::S}{L}{\pc} \frmstep \actframe{S}{L}{\mathit{offset}}}

    \inferrule*
    {\mathit{code}[\pc] = \instr{ifnull}\ \mathit{offset} \\ a \not= \mathit{null}}
    {\actframe{a::S}{L}{\pc} \frmstep \actframe{S}{L}{\pc+1}}

    \inferrule*
    {\mathit{code}[\pc] = \instr{goto}\ \mathit{offset}}
    {\actframe{S}{L}{\pc} \frmstep \actframe{S}{L}{\mathit{offset}}}
  \end{mathpar}  
  \caption{Intra-frame Operational Semantics Rules}
  \label{fig:frm-rules}
\end{figure}

\begin{figure}[t]
  \centering
  \begin{mathpar}
    \inferrule*
    {\mathit{code}[\pc] = \instr{new}\ \mathit{desc} \\
      (\forall \mathit{fnm}.\ (a,\mathit{fnm}) \not\in H)}
    {\actframe{S}{L}{\pc}, H \mutstep \actframe{a::S}{L}{\pc+1}, H[a\mapsto \mathit{desc}], e}

    \inferrule*
    {\mathit{code}[\pc] = \instr{getfield}\ \mathit{fnm} \\
      H[a,\mathit{fnm}] = v}
    {\actframe{a::S}{L}{\pc}, H \mutstep \actframe{v::S}{L}{\pc+1}, H, e}

    \inferrule*
    {\mathit{code}[\pc] = \instr{putfield}\ \mathit{fnm}}
    {\actframe{a::v::S}{L}{\pc}, H \mutstep \actframe{S}{L}{\pc+1}, H[(a,\mathit{fnm})\mapsto v], e}

    \inferrule*
    {\mathit{code}[\pc] = \instr{free}\ \mathit{desc}}
    {\actframe{a::S}{L}{\pc}, H \mutstep \actframe{S}{L}{\pc+1}, H\setminus\langle a,\mathit{desc}\rangle, e}

    \inferrule*
    {\mathit{code}[\pc] = \instr{consume}\ r}
    {\actframe{S}{L}{\pc}, H \mutstep \actframe{S}{L}{\pc+1}, H, r}
  \end{mathpar}
  \caption{Heap and Resource Mutating Operational Semantics Rules}
  \label{fig:mut-rules}
\end{figure}

\begin{figure}[t]
  \centering
  \begin{mathpar}
    \inferrule*
    {f \frmstep f'}
    {\langle r, H, f::\mathit{fs} \rangle \step \langle r, H, f'::\mathit{fs}\rangle}

    \inferrule*
    {f, H \mutstep f', H', r_c}
    {\langle r, H, f::\mathit{fs} \rangle \step \langle r \cdot r_c, H', f'::\mathit{fs}\rangle}

    \inferrule*
    {\mathit{code}[\pc] = \instr{return}}
    {\langle r, H, \actframe{v::S}{L}{\pc}::\langle \mathit{code'}, S', L', \pc' \rangle::\mathit{fs} \rangle \step \langle r, H, \langle \mathit{code'}, v::S', L', \pc' \rangle::\mathit{fs} \rangle}

    \inferrule*
    {\mathit{code}[\pc] = \instr{call}\ \mathit{pname} \\ \mathit{prg}[\mathit{pname}] = \mathit{code'}}
    {\langle r, H, \actframe{\mathit{args}\append S}{L}{\pc}::\mathit{fs}\rangle
      \step
      \langle r, H, \langle \mathit{code'}, [], \ulcorner\mathit{args}\urcorner, 0\rangle::\actframe{S}{L}{\pc+1}::\mathit{fs} \rangle}
  \end{mathpar}
  \caption{Small-step Operational Semantics Rules}
  \label{fig:step-rules}
\end{figure}

The programming language we treat is a simple stack-based virtual
machine, similar to Java bytecode. We have removed the class-based
object system and virtual methods, but retained mutable heap and
procedures.  There are two types: $\mathsf{int}$ and $\mathsf{ref}$,
corresponding to the two kinds of values in $\mathbb{V}$. We assume a
set $\mathbb{P} \ni \mathit{pname}$ of procedure names, where a
procedure's name also determines its list of argument types and its
return type. Programs are organised into a finite set of procedures,
indexed by their name and individually consisting of lists of
instructions from the following collection:
\begin{eqnarray*}
  \iota & ::= & \instr{iconst}\ z \sepbar \instr{ibinop}\ \mathit{op} \sepbar \instr{pop} \sepbar \instr{load}\ n \sepbar \instr{store}\ n \sepbar \instr{aconst\_null} \\
  & & \sepbar \instr{binarycmp}\ \mathit{cmp}\ \mathit{offset} \sepbar \instr{unarycmp}\ \mathit{cmp}\ \mathit{offset} \sepbar \instr{ifnull}\ \mathit{offset} \sepbar \instr{goto}\ \mathit{offset}\\
  & & \sepbar \instr{new}\ \mathit{desc} \sepbar \instr{getfield}\ \mathit{fnm} \sepbar \instr{putfield}\ \mathit{fnm} \sepbar \instr{free}\ \mathit{desc} \sepbar \instr{consume}\ r\\
  & & \sepbar \instr{return} \sepbar \instr{call}\ \mathit{pname}
\end{eqnarray*}
These instructions---apart from $\instr{consume}$---are standard, so
we only briefly explain them. Inside each activation frame, the
virtual machine maintains an operand stack and a collection of local
variables, both of which contain values from the semantic domain
$\mathbb{V}$. Local variables are indexed by natural numbers. The
instructions in the first two lines of the list perform the standard
operations with the operand stack, local variables and program
counter. The third line includes instructions that allocate, free and
manipulate structures stored in the heap. The instruction
$\instr{consume}\ r$ consumes the resource $r$. The $\mathit{desc}$
argument to $\instr{new}$ and $\instr{free}$ describes the structure
to be created on the heap by the fields and their types. The fourth
line contains the procedure call and return instructions that
manipulate the stack of activation frames.

Individual activation frames are tuples $\langle \mathit{code}, S, L,
\mathit{pc} \rangle \in \mathsf{Frm}$ consisting of the list of
instructions from the procedure being executed, the operand stack and
local variables, and the program counter. The first two lines of
instructions that we gave above only operate within a single
activation frame and have no effect on the heap or consumable
resources, so we give their semantics as a small-step relation between
frames: $\frmstep\ \subseteq \mathsf{Frm} \times \mathsf{Frm}$ defined
in \autoref{fig:frm-rules}. The rules make use of semantic
interpretations $\sem{\mathit{op}}$ and $\sem{\mathit{cmp}}$ of binary
operations and binary relations on integers.

The third line of instructions contains those that manipulate the heap
and consume resources. Their small-step operational semantics is
modelled by a relation $\mutstep\ \subseteq \mathsf{Frm} \times
\mathbb{H} \times \mathsf{Frm} \times \mathbb{H} \times \mathcal{R}$,
which relates the before and after activation frames and heaps, and
states the consumable resource consumed by this step. The rules
defining this relation are given in \autoref{fig:mut-rules}. The rules
for the instructions $\instr{new}$ and $\instr{free}$ take a parameter
$\mathit{desc}$. This parameter describes the fields and their types
for the object to be allocated or deallocated. For a description
$\mathit{desc} = \langle \mathit{fnm}_1 : \tau_1, ..., \mathit{fnm}_n
: \tau_n \rangle$, we have written $H[a \mapsto \mathit{desc}]$ to
denote a heap updated with $(a,\mathit{fnm}_i) \mapsto
\mathit{default}(\tau_i)$, where $\mathit{default}(\mathsf{int}) = 0$
and $\mathit{default}(\mathsf{ref}) = \mathit{null}$. For the
$\instr{free}$ instruction, $H\setminus\langle a,
\mathit{desc}\rangle$ denotes the removal of all elements with keys
$(a,\mathit{fnm}_i)$ in $H$.

Note that all the rules in the $\mutstep$ relation apart from the
$\instr{consume}$ instruction consume no resources: $e$ is the
identity element of our resource monoid.

A state of the full virtual machine is a tuple $\langle r, H,
\mathit{fs} \rangle \in \mathsf{State}$, where $r$ is the resource
consumed to this point, $h$ is the current heap, and $\mathit{fs}$ is
a list of activation frames. The small-step operational semantics of
the full machine for some program $\mathit{prg}$ is given by a
relation $\step\ \subseteq \mathsf{State}
\times \mathsf{State}$ which incorporates the $\frmstep$ and
$\mutstep$ relations and also describes how the $\instr{call}$ and
$\instr{return}$ instructions manipulate the stack of activation
frames. The rules defining this relation are presented in
\autoref{fig:step-rules}. In the rule for the instruction
$\instr{call}$, the operation $\append$ denotes list concatenation,
$[]$ denotes the empty list and $\ulcorner-\urcorner$ denotes the
translation of a list to a finite map from natural numbers in the
obvious way.

Finally, we use the predicate $s \downarrow H, r, v$ to indicate when
a $\instr{return}$ instruction is to be executed and there is only one
activation frame on the stack. In this case execution of the program
terminates. The $H, r$ and $v$ are the final heap, the total consumed
resources and the return value of the program respectively.

\subsection{Assertions}\label{sec:logic-assertions}

Every procedure $\mathit{pname}$ in the program is annotated with a
precondition and a post-condition. To allow for variables that are
universally quantified over both the pre- and post-condition we make
use of Coq's dependent types to augment procedure specifications with
a specific ``environment'' type. A procedure specification is a
dependent triple:
\begin{displaymath}
  \langle \mathbb{E} : \mathsf{Type}, P \subseteq \mathbb{E} \times \mathbb{V}^* \times \mathbb{H} \times \mathcal{R}, Q \subseteq \mathbb{E} \times \mathbb{V}^* \times \mathbb{H} \times \mathcal{R} \times \mathbb{V} \rangle
\end{displaymath}

Preconditions $P$ are predicates over $\mathbb{E} \times \mathbb{V}^*
\times \mathbb{H} \times \mathcal{R}$: environments, lists of arguments
to the procedure and the heap and available resource at the start of
the procedure's execution. Post-conditions are predicates over
$\mathbb{E} \times \mathbb{V}^* \times \mathbb{H} \times \mathcal{R}
\times \mathbb{V}$: environments, argument lists and the heap,
remaining consumable resource and return value. 

Intermediate assertions in our program logic are predicates over
$\mathbb{E} \times \mathbb{V}^* \times \mathbb{H} \times \mathcal{R}
\times \mathbb{V}^* \times (\mathbb{N} \rightharpoonup \mathbb{V})$:
environments, argument lists, the heap, remaining consumable resource
and the current operand stack and local variable store. Intermediate
assertions are the assertions that are attached to every instruction
in the body of a procedure by our program logic and specify a
sufficient precondition for safely executing that instruction and all
of its successors.

Note that each of the three different types of assertions talks about
the \emph{remaining} consumable resources available to the program,
not the resources that have already been consumed.

\subsection{Program Logic}

\newcommand{\assn}[5]{#1 \vdash_#2 \left\{#3\right\} \Rightarrow #4\mathord:\mathord#5}

A proof that a given procedure's implementation $\mathit{code}$
matches its specification $\langle \mathbb{E}, P,Q \rangle$ consists
of a map $C$ from instruction offsets in $\mathit{code}$ to assertions
such that:
\begin{enumerate}[(1)]
\item\label{enum:instrs-ok} Every instruction's assertion is suitable
  for that instruction: for every instruction offset $i$ in
  $\mathit{code}$, there exists an assertion $A$ such that
  $\assn{C}{Q}{A}{i}{\mathit{code}[i]}$, and $C[i]$ implies
  $A$. \autoref{fig:rules} gives the definition of the judgement
  $\assn{C}{Q}{A}{i}{\iota}$ for a selected subset of the instructions
  $\iota$. The post-condition $Q$ is used for the case of the
  $\instr{return}$ instruction. We explain the definition of this
  judgement in more detail below.
\item The precondition implies the assertion for the first
  instruction: for all environments $\mathit{env} \in \mathbb{E}$,
  arguments $\mathit{args}$, heaps $H$ and consumable resources $r$,
  we have
  \begin{displaymath}
    P(e, \mathit{args}, H, r) \Rightarrow C[0](\mathit{env}, \mathit{args}, H, r, [], \ulcorner\mathit{args}\urcorner)
  \end{displaymath}
  where $[]$ denotes the empty operand stack, and
  $\ulcorner-\urcorner$ maps lists of values to finite maps from
  naturals to values, as in the operational semantics in
  \autoref{fig:step-rules}.
\end{enumerate}

\noindent When \hyperref[enum:instrs-ok]{condition \ref*{enum:instrs-ok}} holds,
we write this as $C \vdash \mathit{code} : Q$, indicating that the
procedure implementation $\mathit{code}$ has a valid proof $C$ for the
post-condition $Q$.

\begin{figure}[t]
  \centering

  \begin{displaymath}
  \begin{array}{l}
    \assn{C}{Q}{\lambda(\mathit{env}, \mathit{args}, r, H, S, L).\ C[i+1](\mathit{env}, \mathit{args}, r, H, z::S, L)}{i}{\instr{iconst}\ z}
    \\
    \\
    \assn{C}{Q}{\begin{array}{l}
        \lambda(\mathit{env}, \mathit{args}, r, H, S, L). \\
        \quad\forall a, S'. S = a :: S' \Rightarrow  \\
        \quad\quad (a \not= \mathit{null} \Rightarrow C[i+1](\mathit{env}, \mathit{args}, r, H, S', L)) \land \\
        \quad\quad (a = \mathit{null} \Rightarrow C[n](\mathit{env}, \mathit{args}, r, H, S', L)) 
      \end{array}}{i}{\instr{ifnull}\ n}
    \\
    \\
    \assn{C}{Q}{
    \begin{array}{l}
      \lambda(\mathit{env}, \mathit{args}, r, H, S, L). \\
      \quad\forall a, v, S'. \\
      \quad\quad S = a::v::S' \land \\
      \quad\quad (a,\mathit{fnm}) \in H \land \\
      \quad\quad C[i+1](\mathit{env}, \mathit{args}, r, H[(a,\mathit{fnm})\mapsto v], S', L)
    \end{array}}{i}{\instr{putfield}\ \mathit{fnm}}
    \\
    \\
    \assn{C}{Q}{
      \begin{array}{l}
        \lambda(\mathit{env}, \mathit{args}, r, H, S, L). \\
        \quad\exists r'.\ r_c \cdot r' \sqsubseteq r \land C[i+1](\mathit{env}, \mathit{args}, r', H, S, L)
      \end{array}}{i}{\instr{consume}\ r_c}
    \\
    \\
    \assn{C}{Q}{
    \begin{array}{l}
      \lambda(\mathit{env}, \mathit{args}, r, H, S, L). \\
      \quad\forall \mathit{args'}\ \mathit{S'}. S = \mathit{args'} \append S' \Rightarrow \\
      \quad\quad \exists \mathit{env}' \in \mathbb{E}_{\mathit{p}}, (H_1,r_1), (H_2,r_2). \\
      \quad\quad\quad R(H_1,r_1)(H_2,r_2)(H,r) \land \\
      \quad\quad\quad P_{\mathit{p}}(\mathit{env}', \mathit{args'}, H_1, r_1) \land \\
      \quad\quad\quad \forall v, (H'_1, r'_1). \\
      \quad\quad\quad\quad H'_1 \# H_2 \Rightarrow \\
      \quad\quad\quad\quad Q_{\mathit{p}}(\mathit{env}', \mathit{args'}, H'_1, r'_1, v) \Rightarrow \\
      \quad\quad\quad\quad C[i+1](\mathit{env}, \mathit{args'}, r'_1 \cdot r_2, H'_1 \uplus H_2, v::S', L)
    \end{array}}{i}{\instr{call}\ \mathit{p}}
    \\
    \\
    \assn{C}{Q}{\lambda(\mathit{env}, \mathit{args}, r, H, S, L). \forall v, S'.\ S = v::S' \Rightarrow Q(\mathit{env}, \mathit{args}, r, H, v)}{i}{\instr{return}}
  \end{array}
\end{displaymath}
  \caption{Program Logic Rules (Extract)}
  \label{fig:rules}
\end{figure}

The rules in \autoref{fig:rules} are an illustrative subset of the
rules of the program logic. The rule for $\instr{iconst}$ merely
states that the precondition for executing this instruction is the
precondition of the next instruction with the specified integer pushed
on to the stack. The rules for most of the other intra-frame,
non-mutating instructions are similar. Slightly different are the
rules for the conditional instructions, for example $\instr{ifnull}$;
these make use of logical conjunction to make sure that both possible
outcomes of the conditional have their preconditions satisfied.

The rules for $\instr{putfield}$ and $\instr{consume}$ demonstrate
heap and consumable resource consumption respectively. For
$\instr{putfield}$, the heap is judged to be satisfactory if it
contains the address and field that are to be mutated. For resource
consumption, we must ``subtract'' the consumed resource from the
current resource that has been supplied. If the desired resource is
not present then this instruction's precondition will not hold.

The rule for the $\instr{call}$ instruction demonstrates how the heap
and the consumable resources are dealt with similarly for the purposes
of a frame rule. The current heap and consumable resources are split
into two parts $(H_1,r_1)$ and $(H_2,r_2)$, with the first part being
handed off to the callee for it to be mutated. Finally, the
precondition of the $\instr{return}$ instruction is directly derived
from the post-condition of the current procedure.

\subsection{Soundness}\label{sec:program-logic-soundness}

Soundness for the program logic is stated as the preservation of a
safety invariant by every step of the virtual machine. We define
safety for activation frames, frame stacks and whole machine states,
each building on the last.

We say that an activation frame is safe if there is a proof for the
code being executed in the frame such that the requirements of the
next instruction to be executed are satisfied. Formally, a frame $f =
\langle \mathit{code}, S, L, \mathit{pc} \rangle$ is safe for
environment $\mathit{env} \in \mathbb{E}$, arguments $\mathit{args}$,
heap $H$, resource $r$ and post-condition $Q$, written
$\mathit{safeFrame}(\mathit{env}, f, H, r, \mathit{args}, Q)$
if\footnote{In this definition, and all the later ones in this
  section, we have omitted necessary assertions about well-typedness
  of the stack, local variables and the heap because they would only
  clutter our presentation.}:
\begin{enumerate}[(1)]
\item There exists a certificate $C$ such that $C \vdash \mathit{code}
  : Q$;
\item $C[\mathit{pc}]$ exists and $C[\mathit{pc}](\mathit{env},
  \mathit{args}, r, H, S, L)$ holds.
\end{enumerate}

\noindent
Safety of activation frames is preserved by steps in the virtual
machine:

\begin{lem}[Intra-frame safety preservation]\label{lem:frame-safety}\ 

  \begin{enumerate}[\em(1)]
  \item If
    \begin{enumerate}[\em(a)]
    \item $\mathit{safeFrame}(\mathit{env}, f, H, r, \mathit{args},
      Q)$
    \item $f \frmstep f'$
    \end{enumerate}
    then $\mathit{safeFrame}(\mathit{env}, f', H, r, \mathit{args},
    Q)$.
  \item If
    \begin{enumerate}[\em(a)]
    \item $\mathit{safeFrame}(\mathit{env}, f, H_1, r, \mathit{args}, Q)$
    \item $H_1\#H_2$ and $H_1 \uplus H_2 = H$
    \item $f, H \mutstep f', H', r_c$
    \end{enumerate}
    then there exists $H'_1$ and $r'$ such that
    \begin{enumerate}[\em(a)]
    \item $H'_1\#H_2$ and $H'_1 \uplus H_2 = H'$
    \item $r_c \cdot r' \sqsubseteq r$
    \item $\mathit{safeFrame}(\mathit{env}, f', H'_1, r',
      \mathit{args}, Q)$.
    \end{enumerate}
  \end{enumerate}
\end{lem}

\noindent The second part of this lemma states that if we take a step that
mutates the heap or consumes some resource, and the activation frame
has been certified as safe for a sub-part of the heap, then the rest
of the heap---$H_2$---is unaffected by a single step of execution in
this activation frame, and the new state of the activation frame is
safe for the mutated heap and new amount of consumable resources.

\begin{rem}\label{rem:frame-rule}
  We pause for a moment to consider the relationship between our
  program logic and traditional Separation Logic. The second part of
  the previous lemma effectively states that execution steps for
  mutating instructions are \emph{local}: for any other piece of heap
  that is present but not mentioned in its precondition, the execution
  of a mutating instruction will not affect it. This is usually
  expressed in Separation Logic by the frame rule that states if we
  know that $\{P\}C\{Q\}$ holds, then $\{P * R\}C\{Q * R\}$ holds for
  any other resource assertion $R$. We do not have an explicit frame
  rule in our program logic; application of the rule is implicit in
  the rule for the $\instr{call}$ instruction (so, conflatingly, the
  frame rule is applied for new activation frames). We do not have
  access to the frame rule in order to modularly reason about the
  internals of each procedure, e.g. local reasoning about individual
  loops. This is partially a consequence of the unstructured nature of
  the bytecode that we are working with. It has not been a hindrance
  in small examples that we have verified so far, but may well become
  so in larger procedures with multiple loops that need
  invariants---see \autoref{sec:merge-sort} for an example. In such
  cases it may be useful to layer a hierarchical structure, matching
  the loops or other sub-program structure, on top of the unstructured
  bytecode in order to apply frame rules and facilitate local
  reasoning inside procedures.
\end{rem}

We have now handled all the instructions except the $\instr{call}$
and $\instr{return}$ instructions that create and destroy activation
frames. To state soundness of our program logic for these we need to
define what it means for a stack of activation frames to be
safe. Intuitively, a stack of activation frames is a bridge between
the overall arguments $\mathit{args}_{\mathit{top}}$ and
post-condition $Q_{\mathit{top}}$ for the program and the arguments
$\mathit{args}_{\mathit{cur}}$ and post-condition $Q_{\mathit{cur}}$
for the current activation frame, with respect to the current heap $H$
and available consumable resources $r$, such that, when the current
activation frame finishes, its calling frame on the top of the stack
is safe. We write this as
$\mathit{safeStack}_{\mathbb{E}_{\mathit{cur}},
  \mathbb{E}_{\mathit{top}}}(\mathit{fs}, H, r,
\mathit{env}_{\mathit{cur}}, \mathit{args}_{\mathit{cur}},
Q_{\mathit{cur}}, \mathit{env}_{\mathit{top}},
\mathit{args}_{\mathit{top}}, Q_{\mathit{top}})$. The types
$\mathbb{E}_{\mathit{cur}}$ and $\mathbb{E}_{\mathit{top}}$ refer to
the environment types for the current and top-level procedures
respectively, and $\mathit{env}_{\mathit{cur}} \in
\mathbb{E}_{\mathit{cur}}$, $\mathit{env}_{\mathit{top}} \in
\mathbb{E}_{\mathit{top}}$ are the specific elements being used.

Accordingly, we say that the empty frame stack is safe when $r = e$,
$H = \predicate{emp}$, $\mathit{env}_{\mathit{cur}} =
\mathit{env}_{\mathit{top}}$, $\mathit{args}_{\mathit{cur}} =
\mathit{args}_{\mathit{top}}$ and $Q_{\mathit{cur}} =
Q_{\mathit{top}}$. A non-empty frame stack $\mathit{fs} = \langle
\mathit{code}, S, L, \mathit{pc} \rangle::\mathit{fs}'$ is safe when
there exist $(H_1,r_1)$, $(H_2, r_2)$, $\mathit{env}$,
$\mathit{args}$, $Q$ and $C$, $A$ such that:
\begin{enumerate}[(1)]
\item $R(H_1,r_1)(H_2,r_2)(H,r)$;
\item The code is certified: $C \vdash \mathit{code} : Q$;
\item The next instruction to be executed has precondition $A$: $C[\mathit{pc}] = A$;
\item When the callee returns, the instruction's precondition will be
  satisfied: for all $v \in \mathbb{V}, H'_2, r'_2$ such that
  $H'_2\#H_1$ and $Q_{\mathit{cur}}(\mathit{env}_{\mathit{cur}},
  \mathit{args}_{\mathit{cur}}, H'_2, r'_2, v)$ holds,
  $A(\mathit{env}, \mathit{args}, r'_2 \cdot r_1, H'_2 \uplus H_1,
  v::S, L)$ holds as well.
\item The rest of the frame stack $\mathit{fs}$ will be safe when this
  activation frame returns:\\ $\mathit{safeStack}(\mathit{fs}, H_2, r_2,
  \mathit{env}, \mathit{args}, Q, \mathit{env}_{\mathit{top}},
  \mathit{args}_{\mathit{top}}, Q_{\mathit{top}})$.
\end{enumerate}

\noindent Note how the $\mathit{safeStack}$ predicate divides up the heap and
consumable resource between the activation frames on the call stack;
each frame hands a piece of its heap and consumable resource off to
its callees to use. This mirrors the formulation of the rule for
$\instr{call}$ in the program logic in \autoref{fig:rules}.

Finally, we say that a state $s = \langle r_c, H, \mathit{fs} \rangle$
is safe for environment $\mathit{env}$, arguments $\mathit{args}$,
post-condition $Q$ and maximum resource $r_{\mathit{max}}$, written
$\mathit{safeState}(s, \mathit{env}, \mathit{args}, Q,
r_{\mathit{max}})$, if:

\begin{enumerate}[(1)]
\item there exists an $r_{\mathit{future}}$ such that $r_c \cdot
  r_{\mathit{future}} \sqsubseteq r_{\mathit{max}}$; and also
\item $r_{\mathit{future}}$ and $H$ split into $(H_1, r_1)$ and $(H_2,
  r_2)$, i.e. $R(H_1,r_1)(H_2,r_2)(H, r_{\mathit{future}})$;
\item there exists at least one activation frame: $\mathit{fs} = f ::
  \mathit{fs}'$ and environment $\mathit{env}_{\mathit{cur}}$, arguments $\mathit{args}_{\mathit{cur}}$ and
  post-condition $Q_{\mathit{cur}}$; such that
\item $\mathit{safeFrame}(f, H_1, r_1, \mathit{env}_{\mathit{cur}}, \mathit{args}_{\mathit{cur}},
  Q_{\mathit{cur}})$; and
\item $\mathit{safeStack}(\mathit{fs}, H_2,
  r_2,\mathit{env}_{\mathit{cur}}, \mathit{args}_{\mathit{cur}},
  Q_{\mathit{cur}}, \mathit{env}, \mathit{args}, Q)$.
\end{enumerate}

\noindent The key point in the definition of $\mathit{safeState}$ is that the
assertions of the program logic talk about the resources that will be
consumed in the \emph{future} of the program's execution. Safety for a
state says that when we combine the future resource requirements
$r_{\mathit{future}}$ with resources that have been consumed in the
past, $r_c$, then the total is less than the total resources
$r_{\mathit{max}}$ that are allowed for the execution.

\begin{thm}[Soundness]\label{thm:soundness} Assume that all the procedures in
  $\mathit{prg}$ match their specifications.
  \begin{enumerate}[\em(1)]
  \item\label{part:soundness:step} If
    \begin{enumerate}[\em(a)]
    \item $\mathit{safeState}(s, \mathit{env}, \mathit{args}, Q,
      r_{\mathit{max}})$; and
    \item $s \step s'$
    \end{enumerate}
    then $\mathit{safeState}(s', \mathit{env}, \mathit{args}, Q,
    r_{\mathit{max}})$.
  \item If $\mathit{safeState}(s, \mathit{env}, \mathit{args}, Q,
    r_{\mathit{max}})$ and $s \downarrow H, r, v$, then there exists
    an $r'$ such that $Q(\mathit{env}, \mathit{args}, H, r', v)$ and
    $r \cdot r' \sqsubseteq r_{\mathit{max}}$.
  \end{enumerate}
\end{thm}

\noindent In the halting case in this theorem, the existentially quantified
resource $r'$ indicates the resources that the program still had
available at the end of its execution. We are also guaranteed that
when the program halts, the total resource that it has consumed will
be less than the fixed maximum $r_{\mathit{max}}$ that we have set,
and moreover, by \autoref{part:soundness:step} of the theorem, this
bound has been observed at every step of the computation.

\begin{rem}
  Though we assumed it above, the proof of soundness of the program
  logic does not require that the monoid of resources is
  commutative. This opens the way to considering non-commutative
  notions of resource, such as traces. However, constructing a usable
  proof-theory for a mixed commutative/non-commutative notion of
  resource (i.e. heaps and a putative non-commutative consumable
  resource) is hard. Also, the resource acquisition variant of the
  program logic that we describe in the next section requires
  commutativity.
\end{rem}

\subsection{Allowing for Resource Acquisition}\label{sec:acquisition}

The operational semantics and program logic described above assume a
fixed total amount of resource that may be consumed by the
program. The precondition of the main procedure of the program
specifies the amount of resource that will be required for the entire
run. In this section we consider the changes necessary to support
dynamic but fallible acquisition of resources via a special
$\instr{acquire}$ instruction. We provide an example use of this
additional capability in \autoref{sec:acquisition-example}.

We assume that there is a capricious environment that entertains
requests for additional resources from programs. Requests may be
granted or denied. A program may request an additional resource using
a special $\mathsf{acquire}$ instruction. To make things interesting,
and to allow for the example in the
\hyperref[sec:acquisition-example]{next section}, we make the resource
requested by $\mathsf{acquire}$ dynamic and correspondingly modify
$\instr{consume}$, removing its static argument:
\begin{displaymath}
  \iota ::= ... \sepbar \instr{consume} \sepbar \instr{acquire}
\end{displaymath}
For both instructions, we assume that the resource being consumed or
requested is indicated by an integer value on the stack. For the
$\instr{acquire}$ instruction the intended semantics is that if the
request is granted then a $1$ is pushed on to the operand stack,
otherwise a $0$ is pushed.

For the operational semantics, we modify the relation
$\mutstep$ to have an additional ``acquired resources'' component:
\begin{displaymath}
  f, H \mutstep f, H, r_{\mathit{consumed}}, r_{\mathit{acquired}}
\end{displaymath}
For each of the existing heap mutating instructions except
$\instr{consume}$, the acquired resource is equal to the unit of the
resource monoid, $e$.  The operational semantics of $\instr{consume}$
is replaced by the following rule to reflect the dynamic resource
identification:
\begin{displaymath}
  \inferrule*
  { \mathit{code}[\pc] = \instr{consume}}
  { \actframe{z::S}{L}{\pc}, H \mutstep \actframe{S}{L}{\pc+1}, H, \mathrm{res}(z), e }
\end{displaymath}
where we assume the existence of a function $\mathrm{res} : \mathbb{Z}
\to \mathcal{R}$ that names certain consumable resources in
$\mathcal{R}$ by integers. We do not assume that $\mathrm{res}$ is a
monoid homomorphism. Two new operational semantics rules are added for
the $\instr{acquire}$ instruction, for the two possible outcomes of
requesting more resource:
\begin{mathpar}
  \inferrule*
  {\mathit{code}[\pc] = \instr{acquire}}
  {\actframe{z::S}{L}{\pc}, H \mutstep \actframe{1::S}{L}{\pc+1}, H, e, \mathrm{res}(z)}
  
  \inferrule*
  {\mathit{code}[\pc] = \instr{acquire}}
  {\actframe{z::S}{L}{\pc}, H \mutstep \actframe{0::S}{L}{\pc+1}, H, e, e}
\end{mathpar}
States of the virtual machine are modified to be four-tuples $\langle
r_{\mathit{con}}, r_{\mathit{tot}}, H, fs \rangle$ of consumed
resources, total allowed resources, current heap and current
activation frame stack. The invariant that is now to be maintained is
that the $r_{\mathit{con}}$ is always less than or equal to
$r_{\mathit{tot}}$. The only rule from \autoref{fig:step-rules} that
is modified (apart from threading through the unchanged
$r_{\mathit{tot}}$) is the rule incorporating $\mutstep$ into the
relation:
\begin{displaymath}
  \inferrule*
  {f, H \mutstep f', H', r_c, r_a}
  {\langle r, r_{\mathit{tot}}, H, f::\mathit{fs} \rangle \step \langle r \cdot r_c, r_{\mathit{tot}} \cdot r_a, H', f'::\mathit{fs}\rangle}
\end{displaymath}
The halting predicate is extended to a five-place relation $s
\downarrow H, r_{\mathit{con}}, r_{\mathit{tot}}, v$, where
$r_{\mathit{con}}$ is the total consumed resource of the execution and
$r_{\mathit{tot}}$ is the total acquired resource.

Perhaps surprisingly, the assertions (pre- and post-conditions and
instruction level assertions) are left unchanged, as are the rules of
the logic from \autoref{fig:rules} for the original set of
instructions. We modify the rule for resource consumption to take into
account the new dynamic nature:
\begin{displaymath}
  \assn{C}{Q}{
    \begin{array}{l}
      \lambda(\mathit{env},\mathit{args},r,H,S,L).\\
      \quad \forall z, S'.\ S = z::S' \Rightarrow \exists r'. \\
      \quad\quad \mathrm{res}(z) \cdot r' \sqsubseteq r \land C[i+1](\mathit{env},\mathit{args},r',H,S',L)
    \end{array}}{i}{\instr{consume}}
\end{displaymath}
and we add a new rule for resource acquisition:
\begin{displaymath}
  \assn{C}{Q}{\lambda(\mathit{env},\mathit{args}, r, H, S, L).\ 
  \begin{array}{l}
    \forall z, S'.\ S = z::S' \Rightarrow \\
    \quad C[i+1](\mathit{env}, \mathit{args}, r \cdot \mathrm{res}(z), H, 1::S', L) \\
    \quad \mathrel\land C[i+1](\mathit{env}, \mathit{args}, r, H, 0::S', L)
  \end{array}}{i}{\instr{acquire}}
\end{displaymath}
This exactly mirrors the pair of operational semantics rules. An
$\instr{acquire}$ instruction is safe to run if it is safe the execute
the next instruction either with additional available consumable
resource and a $1$ on the stack, or with no additional consumable
resource and a $0$ on the stack.

The definition of safe frame remains unchanged from above, and the
statement of the second item of \hyperref[lem:frame-safety]{Lemma
  \ref*{lem:frame-safety}} is adjusted to account for the possibility
of additional resources being acquired:

\begin{lem}[Intra-frame safety preservation (Resource Acquisition Variant)]\label{lem:frame-safety-2}\ 
  \begin{enumerate}[\em(1)]
  \item[\em(2)] If
    \begin{enumerate}[\em(a)]
    \item $\mathit{safeFrame}(\mathit{env}, f, H_1, r, \mathit{args}, Q)$
    \item $H_1\#H_2$ and $H_1 \uplus H_2 = H$
    \item $f, H \mutstep f', H', r_c, r_a$
    \end{enumerate}
    then there exists $H'_1$ and $r'$ such that
    \begin{enumerate}[\em(a)]
    \item $H'_1\#H_2$ and $H'_1 \uplus H_2 = H'$
    \item $r_c \cdot r' \sqsubseteq r \cdot r_a$
    \item $\mathit{safeFrame}(\mathit{env}, f', H'_1, r',
      \mathit{args}, Q)$.
    \end{enumerate}
  \end{enumerate}
\end{lem}

\noindent The definition of the $\mathit{safeStack}$ predicate remains the same
as before, and the $\mathit{safeState}$ predicate is modified to be of
the form $\mathit{safeState}(s, \mathit{env}, \mathit{args}, Q)$: the
original $r_{\mathit{max}}$ argument is taken to be the
$r_{\mathit{tot}}$ from the state. The key safety property, as before,
is that the consumed resources, plus the future resources, is less
than or equal to the total allowed resources. But now the total
resources allowed may be increased during execution.

The new version of \hyperref[thm:soundness]{Theorem
  \ref*{thm:soundness}} is as follows:

\begin{thm}[Soundness (Resource Acquisition Variant)]\label{thm:soundness-2} Assume that all the procedures in
  $\mathit{prg}$ match their specifications.
  \begin{enumerate}[\em(1)]
  \item If
    \begin{enumerate}[\em(a)]
    \item $\mathit{safeState}(s, \mathit{env}, \mathit{args}, Q)$; and
    \item $s \step s'$
    \end{enumerate}
    then $\mathit{safeState}(s', \mathit{env}, \mathit{args}, Q)$.
  \item If $\mathit{safeState}(s, \mathit{env}, \mathit{args}, Q)$ and
    $s \downarrow H, r_{\mathit{con}}, r_{\mathit{tot}}, v$, then
    there exists an $r'$ such that $Q(\mathit{env}, \mathit{args}, H,
    r', v)$ and $r_{\mathit{con}} \cdot r' \sqsubseteq r_{\mathit{tot}}$.
  \end{enumerate}
\end{thm}

\noindent This theorem gives a comparable guarantee to
\hyperref[thm:soundness]{Theorem \ref*{thm:soundness}}, in that the
(now dynamic) resource bound is respected at every step of the
computation and at the end of execution. By inspection of the
operational semantics, we can see that the resource bound may only be
increased by successful execution of an $\instr{acquire}$ instruction.

\subsection{Resource Acquisition Example}\label{sec:acquisition-example}

To illustrate the utility of the language and logic extended with
resource acquisition, we take the example of \emph{block booking} as
presented by Aspinall, Maier and Stark
\cite{DBLP:conf/fmco/AspinallMS07}. The scenario is that an
application running on a mobile device has a list of telephone numbers
that it wants to send SMS messages to. Permission must be sought from
the user to send these messages, since sending incurs a monetary cost.
It is not necessarily convenient to request permission from the user
when the message is to be sent, so permission is requested in
advance. We use the program logic extended with resource acquisition
to bridge the gap between the acquisition of permission and its
consumption, ensuring that no attempt is made to perform an operation
that has not been authorised. In this scenario, the $\instr{acquire}$
instruction is implemented by actually requesting permission from the
user.

\begin{figure}[t]
\begin{Verbatim}
class PhoneNumberNode {
     public PhoneNumber     number;
     public boolean         permission;
     public PhoneNumberNode next;
}

...

public static void requestPermissions (PhoneNumberNode phoneNumber) {
     while (phoneNumber != null) {
          phoneNumber.permission = acquire (phoneNumber.number);
          phoneNumber = phoneNumber.next;
     }
}
\end{Verbatim}
  
  \caption{Resource Acquisition Example}
  \label{fig:acq-example}
\end{figure}

\autoref{fig:acq-example} shows the code for a method,
\texttt{requestPermissions}, that loops through a list of telephone
numbers, requesting permission for each one. We assume that the calls
to the method \texttt{acquire} are compiled to instances of the
$\instr{acquire}$ instruction. We have taken the liberty of assuming a
value type of telephone numbers rather than re-using the \texttt{int}
type. The result of the resource acquisition attempt is stored in the
\texttt{permission} field of the record for that telephone number.

The specification of the precondition of \texttt{requestPermissions}
is simply that there is a proper linked list on the heap, using the
following inductively defined predicate:
\begin{displaymath}
  \predicate{lseg}(x,y) \equiv (x = y \land \emp) \lor (\exists n,p,z.\ \pointsto{x}{number}{n} * \pointsto{x}{permission}{p} * \pointsto{x}{next}{z} * \predicate{lseg}(z,y))
\end{displaymath}
In the post-condition, we make use of another inductively defined
predicate that states the resources available, dependent on the value
of the \texttt{permission} field:
\begin{displaymath}
  \predicate{lseg}^{\mathsf{send}}(x,y) \equiv
  \begin{array}[t]{l}
    (x = y \land \emp) \\
    \lor\ (\exists n,z.\ \pointsto{x}{number}{n} * \pointsto{x}{permission}{1} * R^{\mathsf{send}(n)} * \pointsto{x}{next}{z} * \predicate{lseg}(z,y)) \\
    \lor\ (\exists n,z.\ \pointsto{x}{number}{n} * \pointsto{x}{permission}{0} * \pointsto{x}{next}{z} * \predicate{lseg}(z,y))
  \end{array}
\end{displaymath}
Thus when the \texttt{permission} field has the value $1$, the list
node is associated with the permission to send to that telephone
number; and when the \texttt{permission} field has the value $0$, no
such associated permission is available. We revisit this kind of
conditionally-resource-carrying predicate in
\autoref{sec:inductive-preds}.

Re-interpreting the $\mathsf{consume}$ instruction as sending an SMS
message, a loop that sends messages to all telephone numbers that the
user has approved now looks much like the simple list iteration
example from the \hyperref[sec:intro]{introduction}, augmented with a
dynamic check to ensure that permission has been sought and received.

\section{Deep Assertion Logic}\label{sec:assertion-logic}

In the \hyperref[sec:logic]{previous section} we described a program
logic but remained agnostic as to the exact form of the assertions
save that they must be predicates over certain domains. This shallow
approach makes the statement and soundness proof easier, but inhibits
discussion of actual specifications and proofs in the logic. In this
section we show how a combination of two variants of the logic of
Bunched Implications (BI) \cite{o'hearn99logic,pym:ptsbi:2002} can be
used to provide a syntax for assertions in our program logic. We
combine boolean BI with affine intuitionistic BI, for describing heaps
and consumable resources respectively.

\subsection{Syntax and Semantics}\label{sec:assertion-logic-synsem}

We made use of three different types of
\hyperref[sec:logic-assertions]{assertion} for the program logic:
procedure pre- and post-conditions, and intermediate assertions within
procedures. These all operate on heaps and consumable resources and
the arguments to the current procedure, but differ in whether they
talk about return values or the operand stack and local variables. To
deal with these differences we assume that we have a set of terms in
our logic, ranged over by $t, t_1, t_2, ...$, that at least includes
logical variables and a constant $\mathit{null}$ for representing the
null reference, and also variables for representing the current
procedure arguments, the return value and the operand stack and local
variables as appropriate.

Formulae are built from at least the following constructors:
\begin{eqnarray*}
  \phi & ::= & \top \sepbar t_1 \bowtie t_2 \sepbar \phi_1 \land \phi_2 \sepbar \phi_1 \lor \phi_2 \sepbar \phi_1 \to \phi_2 \sepbar \predicate{emp} \sepbar \phi_1 * \phi_2 \sepbar \phi_1 \wand \phi_2 \sepbar \forall x. \phi \sepbar \exists x. \phi \\
  & & \sepbar \pointsto{t_1}{f}{t_2} \sepbar R_r \sepbar \dots
\end{eqnarray*}
Where $\bowtie\ \in \{=,\not=\}$.  We can also add inductively defined
predicates as needed, see \autoref{sec:inductive-preds} below. The
only non-standard formula with respect to Separation Logic is $R_r$
which represents the presence of some consumable resource $r$. The
semantics of the assertion logic is given in \autoref{fig:semantics}
as a relation $\models$ between environments and heap/consumable
resource pairs and formulae. We assume a sensible semantics
$\llbracket\cdot\rrbracket_\eta$ for terms in a given
environment.

\begin{figure}
  \centering
\begin{displaymath}
  \begin{array}{@{\eta, x}@{\hspace{0.4em}\models\hspace{0.35em}}l@{\hspace{0.15em}\textrm{iff}\hspace{0.4em}}l}
    \top & \textrm{always} \\
    t_1 \bowtie t_2 & \llbracket t_1 \rrbracket_\eta \bowtie \llbracket t_2 \rrbracket_\eta \\
    \emp & x = (H,r) \textrm{ and }H = \{\} \\
    \pointsto{t_1}{f}{t_2} & x = (H,r) \textrm{ and }H = \{(\llbracket t_1 \rrbracket_\eta, \mathsf{f}) \mapsto \llbracket t_2 \rrbracket_\eta\} \\
    R_{r_i} & x = (H,r) \textrm{ and } r_i \sqsubseteq r \textrm{ and } H = \{\} \\
    \phi_1 \land \phi_2 & \eta, x \models \phi_1 \textrm{ and } \eta, x \models \phi_2 \\
    \phi_1 \lor \phi_2 & \eta, x \models \phi_1 \textrm{ or } \eta, x \models \phi_2 \\
    \phi_1 * \phi_2 & \textrm{exists }y,z. \textrm{ st. }Ryzx\textrm{ and }\eta, y \models\phi_1\textrm{ and }\eta, z\models\phi_2 \\
    \phi_1 \to \phi_2 & \textrm{for all }y. \textrm{ if }x \sqsubseteq y\textrm{ and }\eta, y \models \phi_1 \textrm{ then } \eta, y \models \phi_2 \\
    \phi_1 \wand \phi_2 & \textrm{for all }y, z. \textrm{ if }Rxyz \textrm{ and }\eta, y \models \phi_1\textrm{ then }\eta, z \models \phi_2 \\
    \forall v. \phi & \textrm{for all }a, \eta[v \mapsto a], x \models \phi \\
    \exists v. \phi & \textrm{exists }a, \eta[v \mapsto a], x \models \phi \\
  \end{array}
\end{displaymath}
  \caption{Semantics of assertions}
  \label{fig:semantics}
\end{figure}

As a consequence of having an ordering on consumable resources, and
our chosen semantics of $\predicate{emp}$, $*$ and $\wand$, our logic
contains affine intuitionistic BI as a sub-logic for reasoning purely
about consumable resources.

\begin{prop}
  If $\phi$ is a propositional BI formula with only $R_r$ as atoms,
  then $r \mathrel{\models_{\mathrm{bi}}} \phi$ iff $\eta, (r, h) \models \phi$.
\end{prop}

\subsection{Inductively Defined Shape and Resource Predicates}\label{sec:inductive-preds}

We now present some inductively defined predicates that demonstrate
how heap-resident data structures may have resources associated with
their nodes. We have introduced the \hyperref[sec:intro]{resource-aware
  $\predicate{lseg}$ predicate} that describes a segment of a list
with a resource associated to every element:
\begin{displaymath}
  \predicate{lseg}(r,x,y) \equiv
  (x = y \land \emp) \lor (\exists d, z.\ \pointsto{x}{data}{d} * \pointsto{x}{next}{z} * R^r * \predicate{lseg}(r,z,y))
\end{displaymath}
An alternative that we made use of in the block booking example in
\autoref{sec:acquisition-example} is to only demand resources when the
element data satisfies some predicate, for example when the integer
stored in the node is not equal to zero:
\begin{displaymath}
  \predicate{lseg}^{\not=0}(r,x,y) \equiv
  (x = y \land \emp) \lor (\exists d, z.\ \pointsto{x}{data}{d} * \pointsto{x}{next}{z} * (d \not= 0 \to R^r) * \predicate{lseg}^{\not=0}(r,z,y))
\end{displaymath}
This kind of specification allows the conditional resource property to
be specified locally within the list structure. If we were attempting
explicit resource accounting using sized predicates, we would be
forced to reflect the whole list into the logic and state the resource
requirement in terms of the number of non-zero elements:
\begin{displaymath}
  \predicate{lseg}(l, x, y) \land r = \mathsf{length}(\mathsf{filter}(\lambda x. x \not=0, l))
\end{displaymath}
Reasoning with such global list properties is obviously much harder
than locally reasoning about each individual node as it is processed
by the program.

The amortised approach is not limited to reasoning purely about
singly-linked lists, the standard doubly-linked list and tree
predicates of Separation Logic can be easily augmented with local
resource annotations (we have omitted the data components of these
predicates to save space):
\begin{displaymath}
  \predicate{dlseg}(r,p,x,y) \equiv
  (x = y \land \emp) \lor (\exists z.\ \pointsto{x}{next}{z} * \pointsto{x}{prev}{p} * R^r * \predicate{dlseg}(r,x,z,y))
\end{displaymath}
\begin{displaymath}
  \predicate{tree}(r,x) \equiv
  \begin{array}[t]{l}
    (x = \mathit{null} \land \emp) \\
    \mathrel\lor (\exists y, z.\ \pointsto{x}{left}{y} * \pointsto{x}{right}{z} * R^r * \predicate{tree}(y) * \predicate{tree}(z))
  \end{array}
\end{displaymath}

\noindent These predicates all describe resources that are linear in proportion
to the sizes of the data structures. Using the clever technique of
Hoffmann and Hofmann \cite{hoffmann10amortized}, we can also present
lists with associated resources that are polynomially proportional to
the length of the list, by exploiting the presentation of polynomials
using binomial coefficients. In their system, lists are annotated with
resources that are lists of rational numbers $\langle p_1, ..., p_n
\rangle$. The idea is that a list of length $n$ annotated with such a
list has $\sum_{i=1}^k \binom{n}{i}p_i$ associated resource. We can
give such lists as an inductively defined predicate in our logic:
\begin{displaymath}
  \predicate{lseg}(\overrightarrow{p}, x, z) \equiv (x = z \land \emp) \lor (\exists y.\ \pointsto{x}{next}{y} * R^{p_1} * \predicate{lseg}(\lhd(\overrightarrow{p}), y, z))
\end{displaymath}
where $\lhd(\overrightarrow{p}) = (p_1 + p_2, p_2 + p_3, ..., p_{k-1},
p_k)$ is the additive shift of a resource annotation as defined by
Hoffmann and Hofmann.

\subsection{Heap and Resource
  Separation}\label{sec:resource-separation}

In the logic of \autoref{sec:assertion-logic-synsem}, we only made use
of one kind of separating conjunction, $\phi_1 * \phi_2$, that
separates both heaps and consumable resources. This allows the tight
integration of the heap shapes of various data structures and
consumable resources as shown in the
\hyperref[sec:inductive-preds]{previous section}. Evidently, there are
two other possible combinations that allow sharing of heap or
resources. For example, separation of resources, but sharing of heap:
\begin{displaymath}
  \begin{array}{@{\eta, x}@{\hspace{0.4em}\models\hspace{0.35em}}l@{\hspace{0.15em}\textrm{iff}\hspace{0.4em}}l}
    \phi_1 \stackrel{R}{*} \phi_2 &
    \begin{array}[t]{l}
      x = (H,r)\textrm{ and exists }r_1,r_2. \textrm{ st. }\\
      \quad r_1\cdot r_2\sqsubseteq r \\
      \quad \textrm{and }\eta, (H,r_1) \models\phi_1\textrm{ and }\eta, (H,r_2)\models\phi_2
    \end{array}
  \end{array}
\end{displaymath}
This definition looks like it might be useful to specify that we have
a single data structure on the heap, but two resource views on it. A
need for this kind of situation arose in the merge-sort example in
\autoref{sec:merge-sort}. There, the auxiliary procedure
\texttt{advance}, that advances a pointer a certain number of elements
through a list, was given the specification:
\begin{eqnarray*}
  \textrm{Pre}(\texttt{advance}) & : & \predicate{lseg}(a_0, \texttt{l}, \mathit{null})\\
  \textrm{Post}(\texttt{advance}) & : & \predicate{lseg}(a_0, \texttt{l}, \texttt{retval}) * \predicate{lseg}(a_0, \texttt{retval}, \mathit{null})
\end{eqnarray*}
where $a_0$ was an amount of resource associated with every element of
the list that was to be preserved. Note that \texttt{advance} does not
modify the list in any way and does not consume any
resources. Evidently, this specification is satisfied for any
$a_0$. In the spirit of Separation Logic, we would like to be able to
state the specification of \texttt{advance} without mentioning
resources---because it does not consume or release any---and combine
the specification later on with the fact that the list has some
associated resources. So, we would like to give \texttt{advance} the
specification:
\begin{eqnarray*}
  \textrm{Pre}(\texttt{advance}) & : & \predicate{lseg}(0, \texttt{l}, \mathit{null})\\
  \textrm{Post}(\texttt{advance}) & : & \predicate{lseg}(0, \texttt{l}, \texttt{retval}) * \predicate{lseg}(0, \texttt{retval}, \mathit{null})
\end{eqnarray*}
since it does not require or yield any consumable resource, and then
apply a putative resource-frame rule:
\begin{displaymath}
  \inferrule*
  {\{P\}\ C\ \{Q\}}
  {\{P \stackrel{R}{*} R\}\ C\ \{ Q \stackrel{R}{*} R \}}
\end{displaymath}
Where $R$ would record that every element of the unmodified list has
$a_0$ associated resource. As with the normal frame rule of Separation
Logic, this turns what would be a second-order universally quantified
assertion into an unquantified assertion. This simplification is
crucial for developing automated procedures for discharging
verification conditions as in \autoref{sec:proofsearch}. For the
present example, we can use linear programming to infer the
instantiation of $a_0$, but for general complex composite resources,
or for examples that require polymorphic recursion, this problem could
become much harder.

Unfortunately, in order for this rule to be sound we must ensure that
$C$ does not modify the heap in any way that would violate the
resource associations. There is currently no way to enforce this
within the logic. In the example, this manifests itself as the
inability to guarantee that the list in the post-condition of
\texttt{advance} has exactly the same shape as the list in the
precondition, which would be required to assign a resource to every
element.

\section{Automated Verification}\label{sec:proofsearch}

In this section we describe a verification condition (VC) generation
and proof search procedure for automated verification of programs
against specifications in the program logic, as long as procedures
have been annotated with loop invariants. The restricted subset of
Separation Logic that we use in this section is similar to the subset
used by Berdine et al. \cite{berdine05symbolic}, though instead of
performing a forwards analysis of the program, we generate
verification conditions by backwards analysis and then attempt to
solve them using proof search. We develop our own proof search
procedure rather than re-use an existing Separation Logic-inspired
tool in order to incorporate a key feature of Hofmann and Jost's
amortised system: the use of linear programming to infer resource
annotations \cite{hofmann-jost}. In the system we present here, the
proof search procedure generates linear constraints that can be solved
by linear programming to infer resource annotations. A limitation of
the VC-generate and solve technique we use here is the potential for
exponential blow-up of the verification conditions in the size of
program. The forward symbolic execution approach deals with this by
attempting to prune unreachable paths as soon as possible and merging
similar feasible paths using heuristics.

\subsection{Restricted Assertion Logic}\label{sec:restricted-assertions}

Following Berdine et al., we make use of a highly restricted assertion
logic that forbids arbitrary nesting of the additive and separating
conjunctions and disallows negative occurrences of implications. This
makes proof search practical. For the purposes of resource annotation
inference, consumable resources are represented in the restricted
syntax as linear expressions over a collection of globally
existentially quantified meta-variables. We use $y_1, y_2, $ and so on
for resource variables to be inferred, and $x_1, x_2,$ etc. for logical
variables. Resource variables cannot be quantified over inside
formulae.

The basic assertion of the proof search logic is of the form
\begin{displaymath}
  \bigvee_i \exists \overline{x}.\ \Pi_i \sepbar \Sigma_i \sepbar \Theta_i
\end{displaymath}
where we use the meta-variable $S$ to stand for such assertions.  They
consist of a finite disjunction of clauses, each with a collection of
existentially quantified variables and three collections of
assertions. The first portion, $\Pi$, contains assertions about pure
(non-heap and non-consumable resource) data, which are equalities and
disequalities of the form:
\begin{displaymath}
  P ::= t_1 = t_2 \sepbar t_1 \not= t_2
\end{displaymath}
The terms that we allow in the data and heap assertions are either
variables, or the constant $\mathit{null}$. A collection $\Pi = P_1,
..., P_n$ is interpreted by translation into the logic of
\autoref{sec:assertion-logic} as the additive conjunction of the
$P_i$. The second portion, $\Sigma$, contains assertions about the
heap, which are of the form:
\begin{displaymath}
  X ::= \pointsto{t_1}{f}{t_2} \sepbar \predicate{lseg}(\Theta,t_1,t_2)
\end{displaymath}
Here we have made use of the inductively defined list segment
predicate from \autoref{sec:inductive-preds}.  A collection $\Sigma =
X_1, ..., X_n$ is interpreted as the separating (or multiplicative)
conjunction of the $X_i$. The final portion $\Theta$ is a linear
expression indicating an amount of consumable resource. It is easily
possible to generalise this to multiple resources by considering
multiple named linear expressions. Given a valuation of the resource
meta-variables $y_i$, extended to an interpretation $\llbracket \Theta
\rrbracket$ this is interpreted in the logic of
\autoref{sec:assertion-logic} as the formula $R^{\llbracket \Theta
  \rrbracket r}$ for some fixed resource $r$. A whole composite $\Pi
\sepbar \Sigma \sepbar \Theta$ is interpreted as $\Pi \land (\Sigma *
R^{\llbracket \Theta \rrbracket r})$.

Finally, we have the set of goal formulae that the verification
condition generator will produce and the proof search will solve.
\begin{eqnarray*}
  G & ::= & S * G \sepbar S \wand G \sepbar S \sepbar G_1 \land G_2 \sepbar P \to G \sepbar \forall x. G \sepbar \exists x. G
\end{eqnarray*}
Note that we only allow implications ($\to$ and $\mathord{\wand}$) to
appear in positive positions. This means that we can interpret them in
our proof search as adding extra information to the context.

\subsection{Verification Condition Generation}

Verification condition generation is performed for each procedure
individually by computing weakest liberal preconditions for each
instruction, working backwards from the last instruction in the
method. To resolve loops, we require that the targets of all backwards
jumps have been annotated with loop invariants $S$ that are of the
special form in the \hyperref[sec:restricted-assertions]{previous
  section}. This assumes that the instructions have been sorted into
reverse post-order so we can scan the instructions in reverse order to
collect the verification conditions. We omit the rules that we use for
weakest liberal precondition generation since they are very similar to
the rules for the shallowly embedded logic in \autoref{fig:rules}. The
verification condition generator will always produce a VC for the
entailment of the computed weakest liberal precondition of the first
instruction from the procedure's precondition, plus a VC for each
annotated instruction, being the entailment between the annotation and
the computed weakest liberal precondition. All VCs will have a formula
of the form $S$ as the antecedent and a goal formula $G$ as the
conclusion.

The verification condition generation procedure has been formalised
within the Coq proof assistant and proved sound with respect to the
program logic in \autoref{sec:logic}. By using Coq's module system
\cite{DBLP:conf/tphol/Chrzaszcz03} we have abstracted the verification
condition generator over the particular deep assertion logic used. We
used Coq's program extraction capabilities to extract the verification
condition generator and instantiated it with the proof search logic
described in this section.

\subsection{Proof Search}

The output of the verification condition generation phase is a
collection of problems of the form $S \vdash G$, which can each be
reduced to a finite collection of sequents of the form
$\Pi\sepbar\Sigma\sepbar\Theta \vdash G$. To discharge these proof
obligations, we make use of the I/O interpretation of proof search as
defined for intuitionistic linear logic by Cervesato, Hodas and
Pfenning \cite{cervesato00resource}, along with heuristic rules for
unfolding the inductive list segment predicate. We augment the I/O
model of resource accounting with an additional part that collects
linear constraints that may be fed into a integer linear program
solver, to automatically infer resource annotations.

We use the following judgement form for proof search goals that
collect linear constraints. Here $\mathcal{C}$ is a set of linear
constraints over the resource meta-variables $y_i$:
\begin{displaymath}
  \Pi\sepbar\Sigma\sepbar\Theta \vdash G \outputsep \mathcal{C}
\end{displaymath}
The proof search procedure is defined by the rules shown in
\autoref{fig:goal-driven}, \autoref{fig:matching},
\autoref{fig:contra} and \autoref{fig:list-unfolding}. These rules
make use of several auxiliary judgements:
\begin{center}
  \begin{tabular}{c@{\quad}l}
    $\Pi\sepbar\Sigma\sepbar\Theta \vdash \Sigma_{\mathit{goal}} \outputsep \Sigma_{\mathit{out}}, \Theta_{\mathit{out}}, \mathcal{C}$ & Heap assertion matching \\
    $\Theta \vdash \Theta_{\mathit{goal}} \outputsep \Theta_{\mathit{out}}, \mathcal{C}$                     & Resource matching \\
    $\Pi \vdash \bot$                                                & Contradiction spotting \\
    $\Pi \vdash \Pi'$                                                & Data assertion entailment
  \end{tabular}
\end{center}
The backslash notation used in these judgements follows Cervesato et
al., where in the judgement $\Theta \vdash \Theta_{\mathit{goal}}
\outputsep \Theta_{\mathit{out}}, \mathcal{C}$, the proof context
$\Theta$ denotes the facts used as input and $\Theta_{\mathit{out}}$
denotes the facts that are left over (the output) from proving
$\Theta_{\mathit{goal}}$. The $\mathcal{C}$ component collects the
linear constraints that must hold for the judgement to give a valid
separation logic entailment. A similar interpretation is used for the
heap assertion matching judgement. We do not define the data
entailment or contradiction spotting judgement explicitly here; we
intend that these judgements satisfy the basic axioms of equalities
and disequalities.

The rules in \autoref{fig:goal-driven} are the goal driven search
rules. There is an individual rule for each possible kind of goal
formula. The first two rules are matching rules that match a formula
$S$ against the context, altering the context to remove the heap and
resource assertions that $S$ requires, as dictated by the semantics of
the assertion logic. We must search for a disjunct $i$ that is
satisfied by the current context. There may be multiple such $i$, and
in this case the search may have to backtrack. When the goal is a
formula $S$, then we ask that the left-over heap is empty, in order to
detect memory leaks. Note that the logical variables $x_1,x_2,...$ may
not occur in the constraint sets, so we do not need to handle
universally quantified constraints.

\begin{figure}
  \centering
  \begin{mathpar}
    \inferrule*
    {\textrm{exists } i, \overline{t}.\\
      \Pi\sepbar\Sigma\sepbar\Theta \vdash \Sigma_i[\overline{t}/\overline{x}]\outputsep\Sigma',\Theta',\mathcal{C}_H \\
      \Pi \vdash \Pi_i[\overline{t}/\overline{x}] \\
      \Theta' \vdash \Theta_i[\overline{t}/\overline{x}] \outputsep \Theta'', \mathcal{C}_R \\
      \Pi\sepbar\Sigma'\sepbar\Theta'' \vdash G \outputsep \mathcal{C}_G}
    {\Pi\sepbar\Sigma\sepbar\Theta \vdash \left(\bigvee_i \exists \overline{x}.\ \Pi_i \sepbar \Sigma_i \sepbar \Theta_i\right) * G \outputsep \mathcal{C}_H \cup \mathcal{C}_R \cup \mathcal{C}_G}
    
    \inferrule*
    {\textrm{exists } i, \overline{t}.\\
      \Pi\sepbar\Sigma\sepbar\Theta \vdash \Sigma_i[\overline{t}/\overline{x}]\outputsep\predicate{emp},\Theta', \mathcal{C}_H \\
      \Pi \vdash \Pi_i \\
      \Theta' \vdash \Theta_i \outputsep \Theta'', \mathcal{C}_R}
    {\Pi\sepbar\Sigma\sepbar\Theta \vdash \bigvee_i \exists \overline{x}.\ \Pi_i \sepbar \Sigma_i \sepbar \Theta_i \outputsep \mathcal{C}_H \cup \mathcal{C}_R}
    
    \inferrule*
    {\textrm{forall }i, \overline{x}. \\ \Pi, \Pi_i \sepbar \Sigma, \Sigma_i \sepbar \Theta + \Theta_i \vdash G \outputsep \mathcal{C}}
    {\Pi \sepbar \Sigma \sepbar \Theta \vdash \left(\bigvee_i \exists \overline{x}.\ \Pi_i \sepbar \Sigma_i \sepbar \Theta_i\right) \wand G \outputsep \mathcal{C}}

    \inferrule*
    {\Pi, P\sepbar\Sigma\sepbar\Theta \vdash G \outputsep \mathcal{C}}
    {\Pi\sepbar\Sigma\sepbar\Theta \vdash P \rightarrow G \outputsep \mathcal{C}}

    \inferrule*
    {\Pi\sepbar\Sigma\sepbar\Theta \vdash G_1 \outputsep \mathcal{C}_1 \\ \Pi\sepbar\Sigma\sepbar\Theta \vdash G_2 \outputsep \mathcal{C}_2}
    {\Pi\sepbar\Sigma\sepbar\Theta \vdash G_1 \land G_2 \outputsep \mathcal{C}_1 \cup \mathcal{C}_2}

    \inferrule*
    {\Pi\sepbar\Sigma\sepbar\Theta \vdash G \outputsep \mathcal{C} \\ x \not\in \mathit{fv}(\Pi) \cup \mathit{fv}(\Sigma)}
    {\Pi\sepbar\Sigma\sepbar\Theta \vdash \forall x. G \outputsep \mathcal{C}}

    \inferrule*
    {\Pi\sepbar\Sigma\sepbar\Theta \vdash G[t/x] \outputsep \mathcal{C}}
    {\Pi\sepbar\Sigma\sepbar\Theta \vdash \exists x. G \outputsep \mathcal{C}}
  \end{mathpar}
  \caption{Goal Driven Search Rules}
  \label{fig:goal-driven}
\end{figure}

The matching rules make use of the heap and resource matching
judgements defined in \autoref{fig:matching}. The heap matching
judgements take a data, heap and resource context and attempt to match
a list of heap assertions against them, returning the left over heap,
resources and computed constraints. The first three rules are
straightforward: the empty heap assertion is always matchable,
points-to relations are looked up in the context directly and pairs of
heap assertions are split, threading the contexts through. For the
list segment rules, there are three cases. Either the two pointers
involved in the list are equal, in which case we are immediately done;
or we have a single list cell in the context that matches the start
pointer of the predicate we are trying to satisfy, and we have the
required resources for an element of this list, so we can reduce the
goal by one step; or we have a whole list segment in the context and
we can reduce the goal accordingly. The resource matching rule is
where linear constraints are actually generated; to match a resource
we subtract the desired resource from the available resources and add
a constraint to ensure that there was enough resource to do this. Note
that this rule always succeeds during proof search, but may generate
unsatisfiable constraints. Thus back-tracking may still be required.

\begin{figure}[t]
  \centering
  
  \textbf{Heap Matching Rules:}
  \begin{mathpar}
    \inferrule*
    { }
    {\Pi\sepbar\Sigma\sepbar\Theta \vdash \predicate{emp}\outputsep \Sigma, \Theta, \{\}}

    \inferrule*
    {\Pi \vdash t_1 = t'_1 \\ \Pi \vdash t_2 = t'_2}
    {\Pi\sepbar\Sigma,\pointsto{t_1}{f}{t_2}\sepbar\Theta \vdash \pointsto{t'_1}{f}{t'_2}\outputsep\Sigma, \Theta, \{\}}

    \inferrule*
    {\Pi\sepbar\Sigma\sepbar\Theta \vdash \Sigma_1 \outputsep \Sigma', \Theta', \mathcal{C}_1 \\
      \Pi\sepbar\Sigma'\sepbar\Theta' \vdash \Sigma_2 \outputsep \Sigma'', \Theta'', \mathcal{C}_2}
    {\Pi\sepbar\Sigma\sepbar\Theta \vdash \Sigma_1 * \Sigma_2 \outputsep \Sigma'', \Theta'', \mathcal{C}_1 \cup \mathcal{C}_2}

    \inferrule*
    {\Pi \vdash t_1 = t_2}
    {\Pi\sepbar\Sigma\sepbar\Theta \vdash \predicate{lseg}(\Theta_l,t_1,t_2)\outputsep \Sigma,\Theta, \{ \}}

    \inferrule*
    {\Pi \vdash t_1 = t'_1 \\
      \Theta\vdash \Theta_l\outputsep\Theta', \mathcal{C}_R \\
      \Pi\sepbar\Sigma\sepbar\Theta' \vdash \predicate{lseg}(\Theta_l, t_n,t_2)\outputsep \Sigma',\Theta'', \mathcal{C}_H}
    {\Pi\sepbar\Sigma,\pointsto{t_1}{next}{t_n},\pointsto{t_1}{data}{t_d}\sepbar\Theta\vdash \predicate{lseg}(\Theta_l, t'_1,t_2) \outputsep \Sigma',\Theta'', \mathcal{C}_R \cup \mathcal{C}_H}

    \inferrule*
    {\Pi \vdash t'_1 = t_1 \\
      \Pi\sepbar\Sigma\sepbar\Theta \vdash \predicate{lseg}(\Theta_l,t_2,t_3)\outputsep\Sigma',\Theta', \mathcal{C}}
    {\Pi\sepbar\Sigma,\predicate{lseg}(\Theta_l,t_1,t_2)\sepbar\Theta \vdash \predicate{lseg}(\Theta_l,t'_1,t_3)\outputsep \Sigma', \Theta', \mathcal{C}}
  \end{mathpar}

  \bigskip

  \textbf{Resource Matching Rule:}
  \begin{mathpar}
    \inferrule*
    { }
    {\Theta \vdash \Theta' \outputsep (\Theta - \Theta'), \{ \Theta \geq \Theta'\}}
  \end{mathpar}
  \caption{Matching Rules}
  \label{fig:matching}
\end{figure}

The final two sets of rules operate on the proof search context. The
first set, shown in \autoref{fig:contra}, describe how information
flows from the heap part of the context to the data part. If we know
that two variables both have a points-to relation involving a field
$\mathsf{f}$, then we know that these locations must not be
equal. Similarly, if we know that a variable does point to something,
then it cannot be null. If any contradictions are found using these
rules, then the proof search can terminate immediately for the current
goal. This is provided for by the first rule in \autoref{fig:contra}.

\begin{figure}
  \centering
  \begin{displaymath}
    \inferrule*
    {\Pi \vdash \bot}
    {\Pi\sepbar\Sigma\sepbar\Theta \vdash G \outputsep \{\}}
  \end{displaymath}

  \begin{mathpar}
    \inferrule*
    {\Sigma = \pointsto{t_1}{f}{t}, \pointsto{t_2}{f}{t'}, \Sigma' \\
      \Pi, t_1 \not= t_2\sepbar\Sigma\sepbar\Theta \vdash G \outputsep \mathcal{C}}
    {\Pi\sepbar\Sigma\sepbar\Theta \vdash G \outputsep \mathcal{C}}

    \inferrule*
    {\Sigma = \pointsto{t}{f}{t'}, \Sigma' \\
      \Pi, t \not= \mathit{null}\sepbar\Sigma\sepbar\Theta \vdash G \outputsep \mathcal{C}}
    {\Pi\sepbar\Sigma\sepbar\Theta \vdash G \outputsep \mathcal{C}}
  \end{mathpar}
  \caption{Contradiction Flushing}
  \label{fig:contra}
\end{figure}

The final set of rules performs heuristic unfolding of the inductive
$\predicate{lseg}$ predicate. These rules are shown in
\autoref{fig:list-unfolding}. These rules take information from the
data context and use it to unfold $\predicate{lseg}$ predicates that
occur in the heap context. The first rule is triggered when the proof
search learns that there is a list segment where the head pointer of
the list is not equal to null. In this case, two proof search goals
are produced, one for the case that the list segment is empty and one
for when it is not. The other rules are similar; taking information
from the data context and using it to refine the heap context.

\begin{figure}
  \centering
  \begin{mathpar}
    \inferrule*
    {\Pi \vdash t_1 \not= \mathit{null} \\
      \Pi, t_1 = t_2\sepbar\Sigma\sepbar\Theta \vdash G \outputsep \mathcal{C}_1\\
      \Pi\sepbar\Sigma,\pointsto{t_1}{next}{x},\pointsto{t_1}{data}{y},\predicate{lseg}(R,x,t_2)\sepbar\Theta,R \vdash G \outputsep \mathcal{C}_2}
    {\Pi\sepbar\Sigma,\predicate{lseg}(R,t_1,t_2)\sepbar\Theta \vdash G \outputsep \mathcal{C}_1 \cup \mathcal{C}_2}

    \inferrule*
    {\Pi \vdash t_1 = \mathit{null} \\
      \Pi, t_2 = \mathit{null}\sepbar\Sigma\sepbar\Theta \vdash G \outputsep \mathcal{C}}
    {\Pi\sepbar\Sigma,\predicate{lseg}(R,t_1,t_2)\sepbar\Theta \vdash G \outputsep \mathcal{C}}

    \inferrule*
    {\Pi \vdash t_1 = t_2 \\
      \Pi\sepbar\Sigma\sepbar\Theta \vdash G \outputsep \mathcal{C}}
    {\Pi\sepbar\Sigma, \predicate{lseg}(R,t_1,t_2)\sepbar\Theta \vdash G \outputsep \mathcal{C}}

    \inferrule*
    {\Pi \vdash t_1 \not= t_2 \\
      \Pi\sepbar\Sigma,\pointsto{t_1}{next}{x},\pointsto{t_1}{data}{y},\predicate{lseg}(R,x,t_2)\sepbar\Theta,R \vdash G \outputsep \mathcal{C}}
    {\Pi\sepbar\Sigma,\predicate{lseg}(R,t_1,t_2)\sepbar\Theta \vdash G \outputsep \mathcal{C}}
  \end{mathpar}

  \caption{List Unfolding Rules}
  \label{fig:list-unfolding}
\end{figure}

The proof search strategy that we employ works by first saturating the
context by repeatedly applying the rules in \autoref{fig:contra} and
\autoref{fig:list-unfolding} to move information from the data context
into the heap context and vice versa. This process terminates because
there are a finite number of points-to relations and list segment
predicates to generate rule applications, and when new predicates are
introduced via list segment unfolding they either do not trigger any
new inequalities or are over fresh variables about which nothing is
yet known. Once the context is fully saturated, the proof search
reduces the goal by using the goal-driven search rules and the process
begins again.

Given a collection of verification conditions and a successful proof
search over them that has generated a set of linear constraints, we
input these into a linear solver, along with the constraint that every
variable is positive and an objective function that attempts to
minimise variables appearing in the precondition.

\begin{thm}
  The proof search procedure is terminating. Moreover, it is sound: if
  $\Pi \sepbar \Sigma \sepbar \Theta \vdash G \outputsep \mathcal{C}$
  and there is an valuation of the $\overline{y}$ that satisfies
  $\mathcal{C}$, then $\Pi \land (\Sigma * R^{\llbracket \Theta
    \rrbracket r}) \vdash G$ under this valuation using the
  translation of the proof search logic into the logic of
  \autoref{sec:assertion-logic} as defined in
  \autoref{sec:restricted-assertions}.
\end{thm}

\section{Examples of Automated Verification}\label{sec:auto-examples}

\subsection{Small Examples}

We have tested the automated resource inference procedure described in
the \hyperref[sec:proofsearch]{previous section} on several small
examples, including the simple loop iteration example from the
\hyperref[sec:intro]{introduction} and the examples from
\autoref{sec:examples}. We summarise these examples in the following
table. Note that the proof search procedure also verifies the memory
safety of these examples. For timings, these tests were performed on
an PC running 64-bit Ubuntu Linux 10.10 on an 8-core Intel Core i7 860
at 2.80GHz. Times were measured using the GNU \texttt{time}
utility. Our implementation only makes use of a single core.

\medskip

\begin{center}
  \begin{tabular}{|l|p{8cm}|l|}
    \hline
    Name & Property Inferred & Time (s) \\
    \hline
    \texttt{iterate\_list} & Number of iterations is length of input list & 0.010 \\
    \texttt{iterate\_recursive} & Number of calls is length of input list & 0.010 \\
    \texttt{copy\_list}    & Number of allocations is length of input list & 0.013 \\
    \texttt{reverse}       & Number of iterations is length of input list & 0.012 \\
    \texttt{queue}         & Resource annotations on \texttt{enqueue} and \texttt{deqeueue} & 0.019 \\
    \texttt{frying\_pan}   & Number of times each element is visited & 0.018 \\
    \texttt{mergesort}     & Number of comparisons is length of input list & 0.037 \\
    \texttt{tree\_traverse} & Number of calls is size of input tree & 0.012 \\
    \texttt{tree\_copy}    & Number of allocations is size of input tree & 0.014 \\
    \texttt{tree\_mirror}  & Number of calls is size of input tree & 0.012 \\
    \hline
  \end{tabular}
\end{center}

\medskip

In each of these examples, we seeded the program with loop invariants
describing the shape of heap data structures, but left our
implementation to infer the resource bounds. The last three examples
involving trees make use of the inductive $\predicate{tree}$ predicate
from \autoref{sec:inductive-preds}. The proof search procedure was
augmented with heuristic rules for matching and unfolding instances of
the $\predicate{tree}$ predicate, following those for lists. The tree
examples are all recursive procedures.

As can be seen from the table, the time taken for each of the examples
is trivial. It remains to be seen how well this technique scales to
real-world code. It seems evident that the VC-generate and solve
process is not scalable in general due to the potential for
exponential blow-up in the size of the generated formulae. A more
realistic implementation of the program logic described in this paper
would likely use the forward symbolic execution approach, as described
by Berdine et al. \cite{berdine05symbolic}.

\subsection{Frying Pan List Reversal}\label{sec:fryingpan}

As a larger example, we demonstrate the use of the proof search
procedure coupled with linear constraint generation on the standard
imperative in-place list reversal algorithm on lists with cyclic tails
(also known as ``frying pan'' lists). This example was used by
Brotherston, Bornat and Calcagno \cite{brotherston08cyclic} to
illustrate the use of cyclic proofs to prove program termination. Here
we show how our amortised resource logic can be used to infer bounds
on the time complexity of this procedure.
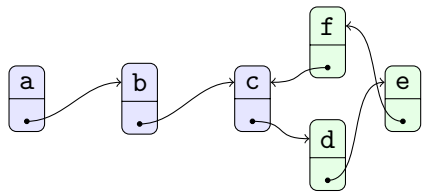
\begin{center}
  \begin{tikzpicture}
    \foreach \n / \x in {a/0, b/1.5, c/3} {
        \node (\n) [listnode] at (\x,0) {\n};
    }

    \node(d) [listnode2] at (4, -0.75) {d};
    \node(e) [listnode2] at (5, 0) {e};
    \node(f) [listnode2] at (4, 0.75) {f};

    \foreach \a / \b in {a/b, b/c, c/d, d/e} \link{\a}{\b};
    \foreach \a / \b in {e/f, f/c} \linkback{\a}{\b};
  \end{tikzpicture}
\end{center}
The ``handle'' of the structure consists of the nodes \texttt{a},
\texttt{b}, \texttt{c} and the ``pan'' consists of the nodes
\texttt{d}, \texttt{e} and \texttt{f}. When the in-place list-reversal
procedure is run upon a structure of this shape, it will proceed up
the handle, reversing it, around the pan, reversing it, and then back
down the handle, restoring it to its original order. For the purposes
of this example, we assume that it takes one element of resource to
handle the reversal of one node. Following Brotherston, Bornat and
Calcagno, we can specify a cyclic list in Separation Logic by the
following formula, where $v_0$ points to the head of the list and
$v_1$ points to the join between the handle and the pan\footnote{Note
  that the $\mathsf{data}$ part of the list node has been omitted in
  this formula and in the loop invariants to reduce clutter.}.
\begin{displaymath}
  \exists k. \predicate{lseg}(x_1, v_0, v_1) * \pointsto{v_1}{next}{k} * \predicate{lseg}(x_2,k,v_1) * R^{x_3}
\end{displaymath}
We have annotated the list segments involved with resource annotation
variables $x_1$ and $x_2$ that we will instantiate using linear
programming. The predicate $R^{x_3}$ denotes any extra resource we may
require. Similarly, we have annotated the required loop invariant
(adapted from Brotherston et al.):
\begin{displaymath}
  \begin{array}{l}
    (\exists k.\ \predicate{lseg}(a_1, l_0, v_1) * \predicate{lseg}(a_2, l_1, \mathit{null}) * \pointsto{v_1}{next}{k} * \predicate{lseg}(a_3, k, v_1) * R^{a_4}) \\
    \lor\ (\exists k.\ \predicate{lseg}(b_1, k, \mathit{null}) * \pointsto{j}{next}{k} * \predicate{lseg}(b_2, l_0, v_1) * \predicate{lseg}(b_3, l_1, j) * R^{b_4}) \\
    \lor\ (\exists k.\ \predicate{lseg}(c_1, l_0, \mathit{null}) * \predicate{lseg}(c_2, l_1, v_1) * \pointsto{v_1}{next}{k} * \predicate{lseg}(c_3, k, v_1) * R^{c_4})
  \end{array}
\end{displaymath}
Each disjunct of the loop invariant corresponds to a different phase
of the procedure's progress. Brotherston et al. note that it is
possible to infer the shape part of this loop invariant using current
Separation Logic tools. Here, we have added the ability to infer
resource annotations, and hence bounds on the time consumption of the
procedure. Running our tool on this example produces the following
instantiation of the variables:

\begin{tabular}{l@{\quad}l@{\quad}l@{\quad}l@{\quad}l}
  Precondition & $x_1 = 2$ & $x_2 = 1$ & $x_3 = 2$ & \\
  Loop invariant, phase 1 & $a_1 = 2$ & $a_2 = 1$ & $a_3 = 1$ & $a_4 = 2$ \\
  Loop invariant, phase 2 & $b_1 = 1$ & $b_2 = 1$ & $b_3 = 0$ & $b_4 = 1$ \\
  Loop invariant, phase 3 & $c_1 = 1$ & $c_2 = 0$ & $c_3 = 0$ & $c_4 = 0$ \\
  Post-condition & $x'_1 = 0$ & $x'_2 = 0$ & $x'_3 = 0$ &
\end{tabular}

\noindent Pictorially, the inference has associated the following amount of
resource with each part of the input structure:
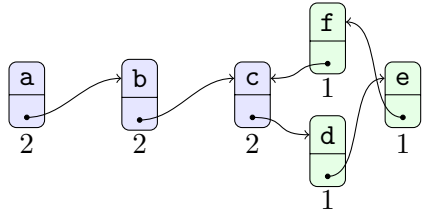
\begin{center}
  \begin{tikzpicture}
    \foreach \n / \x in {a/0, b/1.5, c/3} {
        \node (\n) [listnode] at (\x,0) {\n};
        \node at (\x,-0.65) {2};
    }

    \foreach \n / \x / \y in {d/4/-0.75, e/5/0, f/4/0.75} {
        \node (\n) [listnode2] at (\x,\y) {\n};
        \node at (\x,\y-0.65) {1};
    }

    \foreach \a / \b in {a/b, b/c, c/d, d/e} \link{\a}{\b};
    \foreach \a / \b in {e/f, f/c} \linkback{\a}{\b};
  \end{tikzpicture}
\end{center}
Each node of the handle has 2 associated elements of resource, to
handle the two passes of the handle that the procedure takes, while
the pan has one element of resource for each node. The inferred
annotations for the loop invariant track how the resources on each
node are consumed by the procedure, gradually all reducing to
zero. Since we have added a $\instr{consume}$ instruction to be
executed every time the procedure starts a loop, the resource
inference process has also verified the termination of this procedure,
and given us a bound on the number of times the loop will execute in
terms of the shape of the input.

\section{Conclusions}

We have presented a program logic that extends the resource reasoning
capabilities of Separation Logic from reasoning about mutable
resources such as the heap to consumable resources such as time. We have
demonstrated how doing so allows tight connections between the shape
of data structures and the resources required to process them to be
stated, and so expanding the reach of Separation Logic's local
reasoning principle to consumable resources.

We have presented an automated proof procedure that takes programs
annotated with shape invariants and infers consumable resource
annotations. The main limitation of this automated proof search
procedure is that it only supports the statement and inference of
bounds that are linear in the size of lists that are mentioned in a
procedure's precondition. This is a limitation shared with the
original work of Hofmann and Jost \cite{hofmann-jost}. We note that
this is not a limitation of the program logic that we have presented,
only of the automated verification procedure that we have layered on
top. In \autoref{sec:inductive-preds} we presented a Separation Logic
version of the polynomial potential lists of Hoffmann and Hofmann
\cite{hoffmann10amortized}, which opens the way to inference of
polynomial bounds for pointer manipulating list programs. Initial
experiments with extending our implementation in this direction have
been promising.

We have demonstrated that the use of mixed shape and resource
assertions can simplify the complexity of specifications that talk
about resources, and this should extend to extensions of the proof
search procedure, or to interactive systems based on this program
logic. The resource aware program logic of Aspinall et al.
\cite{aspinall07program} also uses the same layering: a general
program logic for resources (which is proved complete in their case)
is used as a base for a specialised logic for reasoning about the
output of the Hofmann-Jost system.

A possible direction for future work is to consider different
assertion logics and their expressiveness in terms of the magnitude of
resources they can express. We conjecture that the deep assertion
logic we have presented here, extended with the $\predicate{lseg}$
predicate can express resources linear in the size of the heap. It
would be interesting to consider more expressive logics and evaluate
them from the point of view of implicit computational complexity; the
amount of resource that one can express in an assertion dictates the
amount of resource that is available for the future execution of the
program.

Additional future work is to consider the proof theory of the combined
Boolean BI and affine intuitionistic BI that have used in this paper.

Other resource inference procedures that are able to deal with
non-linear bounds include those of Chin et al.
\cite{chin08analysing,chin05memory}, Albert et
al. \cite{albert07costa} and Gulwani et
al. \cite{gulwani09speed}. When dealing with heap-based data
structures, all of these techniques use a method of attaching size
information to assertions about data structures. As we demonstrated in
\autoref{sec:func-queues}, this can lead to unwanted additional
complexity in specifications. However, all of these techniques deal
with numerically bounded loops, which our current prototype automated
procedure cannot. We are currently investigating how to extend our
approach to deal with non-linear and numerically-driven resource
bounds.

\medskip

\noindent
\textbf{Acknowledgements} I would like to thank Kenneth MacKenzie and
Brian Campbell for discussion and comments on this work. The ESOP 2010
and LMCS anonymous reviewers also provided helpful suggestions. This
work was funded by EPSRC Follow-on Fund grant EP/G006032/1 ``Resource
Static Analysis'' and EPSRC grant EP/G068917/1 ``Categorical
Foundations for Indexed Programming''.

\bibliography{amortised} \bibliographystyle{plain}

\end{document}